\documentclass[11pt,twoside,a4paper,cmspaper,final,collab]{cms-tdr}
\def\svnVersion{9c21315}\def\svnDate{2026/07/08}\def\cmsCernNoTag{CERN-EP-2025-270}\def\cmsCernDate{\today}\def\cmsMessage{Published in the Journal of High Energy Physics as \href{http://dx.doi.org/10.1007/JHEP07(2026)005}{\doi{10.1007/JHEP07(2026)005}.}}
\begin{document}\cmsNoteHeader{B2G-25-003}

\newcommand{\pp}{\ensuremath{\Pp\Pp}\xspace}
\newcommand{\SD}{\ensuremath{\text{SD}}\xspace}
\providecommand{\cmsTable}[1]{\resizebox{\textwidth}{!}{#1}}
\newlength\cmsTabSkip\setlength{\cmsTabSkip}{1ex}

\newcommand{\cF}{\ensuremath{c_{\mathrm{F}}}\xspace}
\newcommand{\gF}{\ensuremath{g_{\mathrm{F}}}\xspace}
\newcommand{\gV}{\ensuremath{g_{\PV}}\xspace}
\newcommand{\cH}{\ensuremath{c_{\PH}}\xspace}
\newcommand{\gH}{\ensuremath{g_{\PH}}\xspace}
\newcommand{\gq}{\ensuremath{g_{\mathrm{q}}}\xspace}
\newcommand{\gql}{\ensuremath{g_{\mathrm{q}_{12}}}\xspace}
\newcommand{\gqh}{\ensuremath{g_{\mathrm{q}_{3}}}\xspace}
\newcommand{\gl}{\ensuremath{g_{\mathrm{\Pell}}}\xspace}
\newcommand{\gll}{\ensuremath{g_{\mathrm{\Pell}_{12}}}\xspace}
\newcommand{\glh}{\ensuremath{g_{\mathrm{\Pell}_{3}}}\xspace}
\newcommand{\sigmahat}{\ensuremath{\hat{\sigma}}\xspace}

\DeclareRobustCommand{\qbarpr}{{\HepParticle{\PAQq}{}{(\prime)}}\Xspace}
\newcommand{\taunu}{\ensuremath{\PGt\PAGn}\xspace}
\newcommand{\lnu}{\ensuremath{\Pell\PAGn}\xspace}
\newcommand{\ellell}{\ensuremath{\Pellp\Pellm}\xspace}
\newcommand{\tautau}{\ensuremath{\PGtp\PGtm}\xspace}
\newcommand{\tb}{\ensuremath{\PQt\PAQb}\xspace}
\newcommand{\qqbarprime}{\ensuremath{\PQq\qbarpr}\xspace}
\newcommand{\jet}{\ensuremath{j}}
\newcommand{\jetb}{\ensuremath{j^{\PQb}}}
\newcommand{\jetV}{\ensuremath{j^{\PV}}}
\newcommand{\jettop}{\ensuremath{j^{\PQt}}}
\newcommand{\jetbb}{\ensuremath{j^{\PQb\PQb}}}
\newcommand{\jetH}{\ensuremath{j^{\PH}}}

\newcommand{\Zprime}{\PZpr}
\newcommand{\Wprime}{\PWpr}
\newcommand{\Wprimepm}{{\HepParticle{\PW}{}{\prime\pm}}\Xspace}
\newcommand{\PVpr}{{\HepParticle{\PV}{}{\prime}}\Xspace}
\newcommand{\Vprime}{\PVpr}

\newcommand{\ZH}{\ensuremath{\PZ\PH}\xspace}
\newcommand{\WH}{\ensuremath{\PW\PH}\xspace}
\newcommand{\VH}{\ensuremath{\PV\PH}\xspace}
\newcommand{\WW}{\ensuremath{\PW\PW}\xspace}
\newcommand{\WZ}{\ensuremath{\PW\PZ}\xspace}
\newcommand{\WV}{\ensuremath{\PW\PV}\xspace}
\newcommand{\ZZ}{\ensuremath{\PZ\PZ}\xspace}
\newcommand{\VV}{\ensuremath{\PV\PV}\xspace}
\newcommand{\ZprimeZH}{\ensuremath{\Zprime\to\ZH}\xspace}
\newcommand{\WprimeWZ}{\ensuremath{\Wprime\to\WZ}\xspace}
\newcommand{\VprimeWVWH}{\ensuremath{\Vprime\to\WV/\WH}\xspace}
\newcommand{\VprimeVVVH}{\ensuremath{\Vprime\to\VV/\VH}\xspace}

\newcommand{\Wprimelnu}{\ensuremath{\Wprime\to\lnu}\xspace}
\newcommand{\Wprimetaunu}{\ensuremath{\Wprime\to\taunu}\xspace}
\newcommand{\Wprimetb}{\ensuremath{\Wprime\to\tb}\xspace}
\newcommand{\Zprimett}{\ensuremath{\Zprime\to\ttbar}\xspace}
\newcommand{\Zprimebb}{\ensuremath{\Zprime\to\bbbar}\xspace}
\newcommand{\Zprimetautau}{\ensuremath{\Zprime\to\tautau}\xspace}
\newcommand{\Zprimell}{\ensuremath{\Zprime\to\ellell}\xspace}
\newcommand{\Vprimeqq}{\ensuremath{\Vprime\to\qqbarprime}\xspace}
\newcommand{\Vprimequarks}{\ensuremath{\Vprime\to\text{quarks}}\xspace}
\newcommand{\Vprimeleptons}{\ensuremath{\Vprime\to\text{leptons}}\xspace}
\newcommand{\Vprimebosons}{\ensuremath{\Vprime\to\text{bosons}}\xspace}

\newcommand{\mV}{\ensuremath{m_{\Vprime}}\xspace}
\newcommand{\mZpr}{\ensuremath{m_{\PZpr}}\xspace}
\newcommand{\mWpr}{\ensuremath{m_{\PWpr}}\xspace}
\newcommand{\mW}{\ensuremath{m_{\PW}}\xspace}
\newcommand{\btagged} {\ensuremath{\PQb}\text{-tagged}\xspace}
\newcommand{\bbtagged} {\ensuremath{\text{double-}\PQb}\text{ tagged}\xspace}
\newcommand{\ttagged} {\ensuremath{\PQt}\text{ tagged}\xspace}
\newcommand{\Vtagged} {\ensuremath{\PV}\text{ tagged}\xspace}
\newcommand{\Htagged} {\ensuremath{\PH}\text{ tagged}\xspace}

\newcommand{\Zjets}{\PZ{}+jets\xspace}
\newcommand{\Gjets}{\PGg{}+jets\xspace}

\newcommand{\SI}{S}
\newcommand{\SP}{Sh}
\newcommand{\BA}{B}
\newcommand{\SB}{SB}
\newcommand{\UC}{U}

\cmsNoteHeader{B2G-25-003}
\title{Combination of searches for heavy vector boson resonances in proton-proton collisions at \texorpdfstring{$\sqrt{s} = 13\TeV$}{sqrt(s) = 13 TeV}}

\author[cern]{The CMS Collaboration}
\date{\today}
\abstract{A combined statistical analysis of searches for heavy vector boson resonances decaying into pairs of \PW, \PZ, or Higgs bosons, as well as into quark pairs (\qqbar, \bbbar, \ttbar, \tb) or lepton pairs (\ellell, \lnu), with $\Pell = \Pe, \PGm, \PGt$, is presented. The results are based on proton-proton collision data at a center-of-mass energy of 13\TeV, corresponding to an integrated luminosity of 138\fbinv, collected by the CMS experiment from 2016 to 2018. No significant deviation from the expectations of the standard model is observed. The results are interpreted in the simplified heavy vector triplet (HVT) framework, setting 95\% confidence level upper limits on the production cross sections and on the coupling strengths of the HVT bosons to standard model particles. The results exclude HVT resonances with masses below 5.5\TeV in a weakly coupled scenario, below 4.8\TeV in a strongly coupled scenario, and up to 2.0\TeV in the case of production via vector boson fusion. The combination provides the most stringent constraints to date on new phenomena predicted by the HVT model.}

\hypersetup{
pdfauthor={CMS Collaboration},
pdftitle={Combination of CMS searches for heavy vector boson resonances at sqrt(s)= 13 TeV},
pdfsubject={CMS},
pdfkeywords={CMS, combination, heavy vector triplet, HVT}}

\maketitle

\section{Introduction}\label{sec:Intro}

Over the past half century, searches at particle colliders for heavy resonances decaying into two-body final states have led to several discoveries.
For example, the discovery of the \PW and \PZ bosons in the 1980s through their leptonic decay channels provided critical evidence compatible with the standard model (SM)~\cite{UA1:1983crd,UA2:1983tsx,UA1:1983mne,UA2:1983mlz} predictions.
More recently, the observation of the Higgs (\PH) boson in 2012, decaying into both bosonic and fermionic channels,
confirmed the electroweak symmetry breaking mechanism~\cite{PRLB2012-716-1-1,CMS-PAPERS-HIG-12-028,Chatrchyan:1529865}.

The search for heavy resonances is a central component of the physics program at the CERN LHC,
aiming to uncover physics beyond the SM (BSM) across a wide range of final states~\cite{del_Aguila_2010,de_Blas_2013,CMS:2019kaf,ATLAS:2024qvg}.
Among the theoretized BSM resonances, the spin-1 heavy partners of the \PW and \PZ bosons are named \Wprime and \Zprime, respectively, and are collectively referred to as \Vprime.
At the LHC, both the ATLAS and CMS Collaborations have conducted extensive searches for \Vprime bosons that couple to SM particles via electroweak interactions~\cite{ATLAS:2024qvg,CMS:2019kaf,CMS:2024phk}.
These analyses search for the decay of a \Vprime boson into two massive SM bosons (\PW, \PZ, or \PH), two leptons, or two quarks.

\begin{figure}[ht]
\centering
\includegraphics[width=0.3\textwidth]{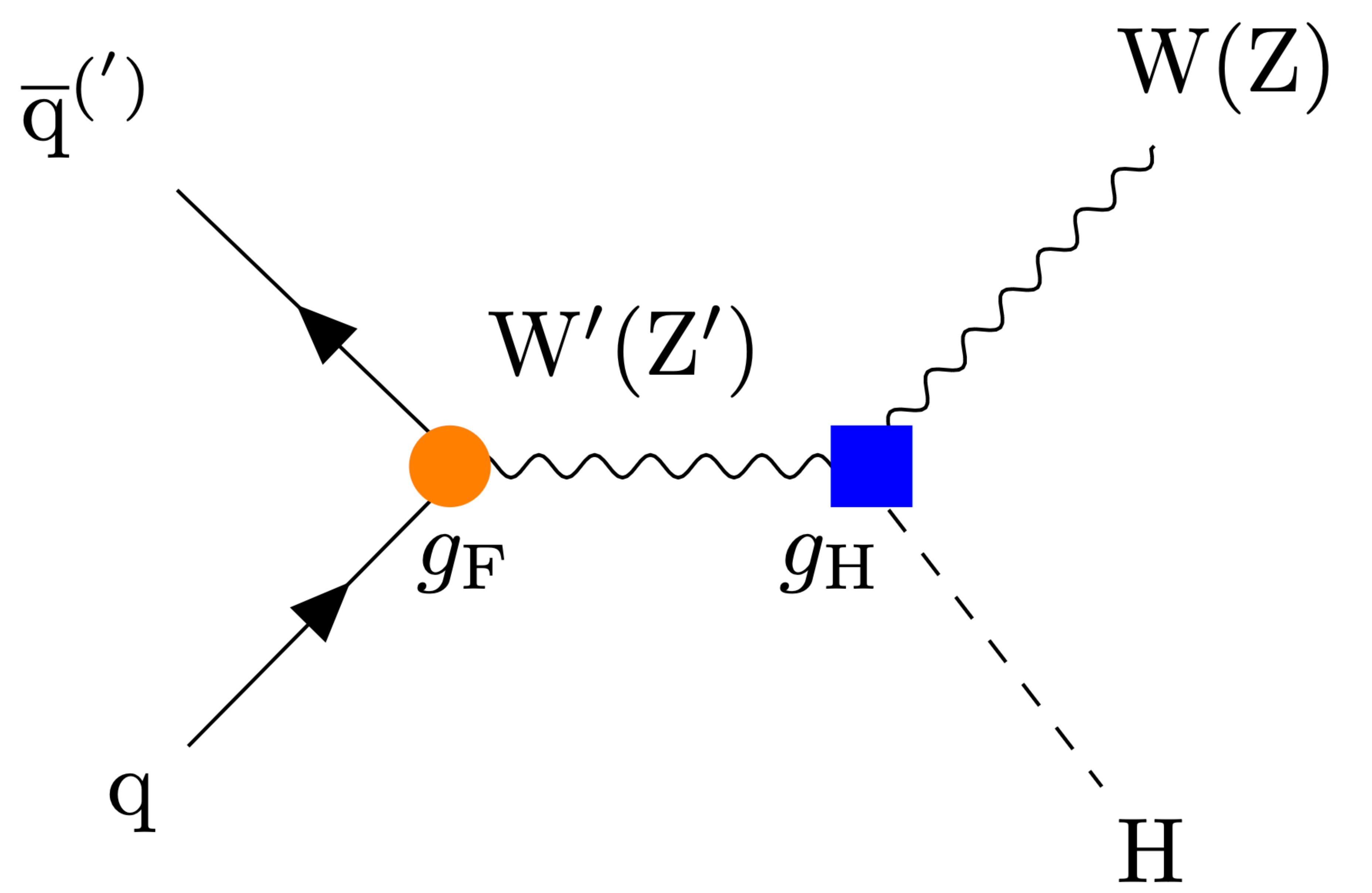}
\includegraphics[width=0.3\textwidth]{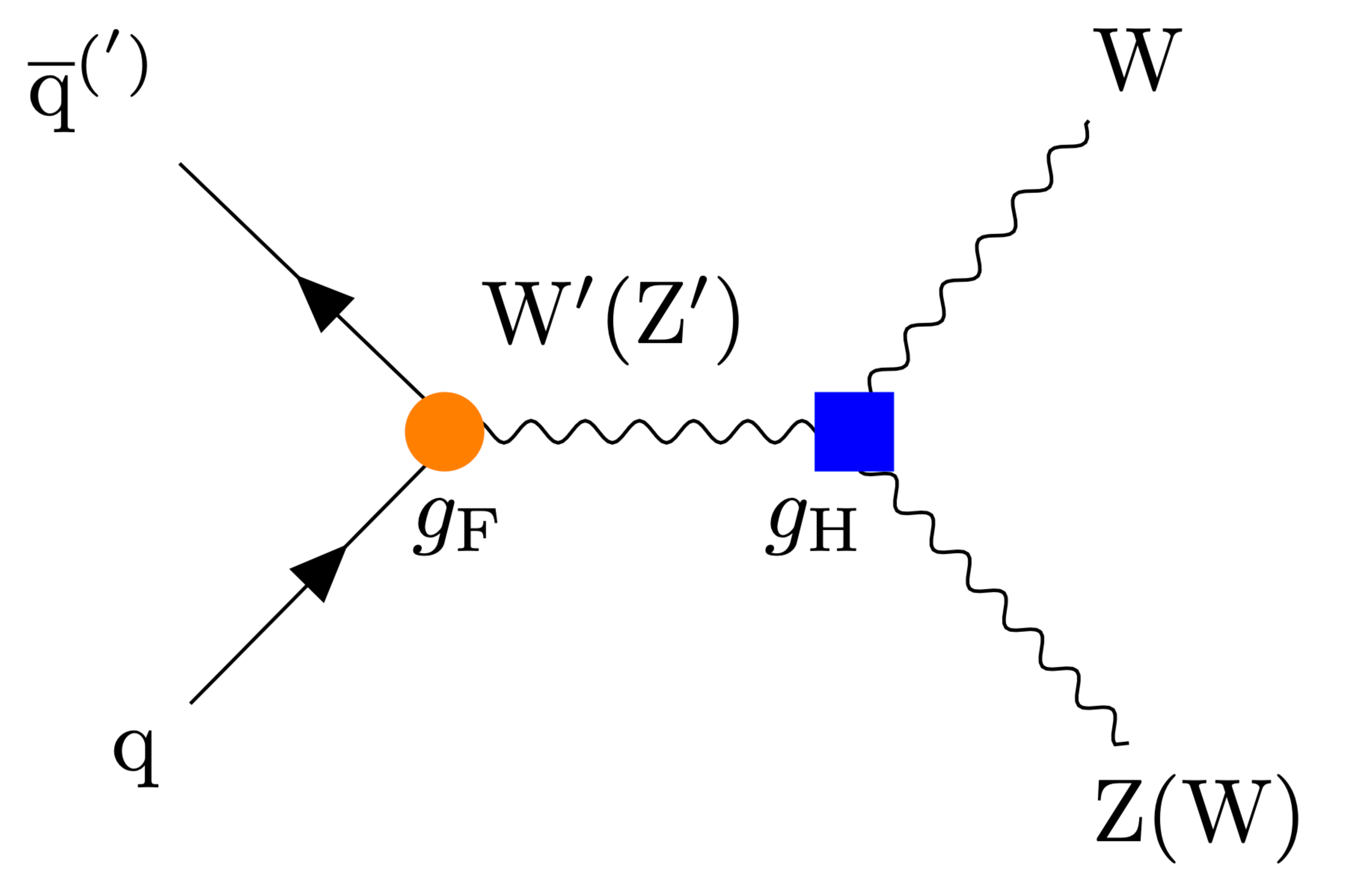}
\includegraphics[width=0.3\textwidth]{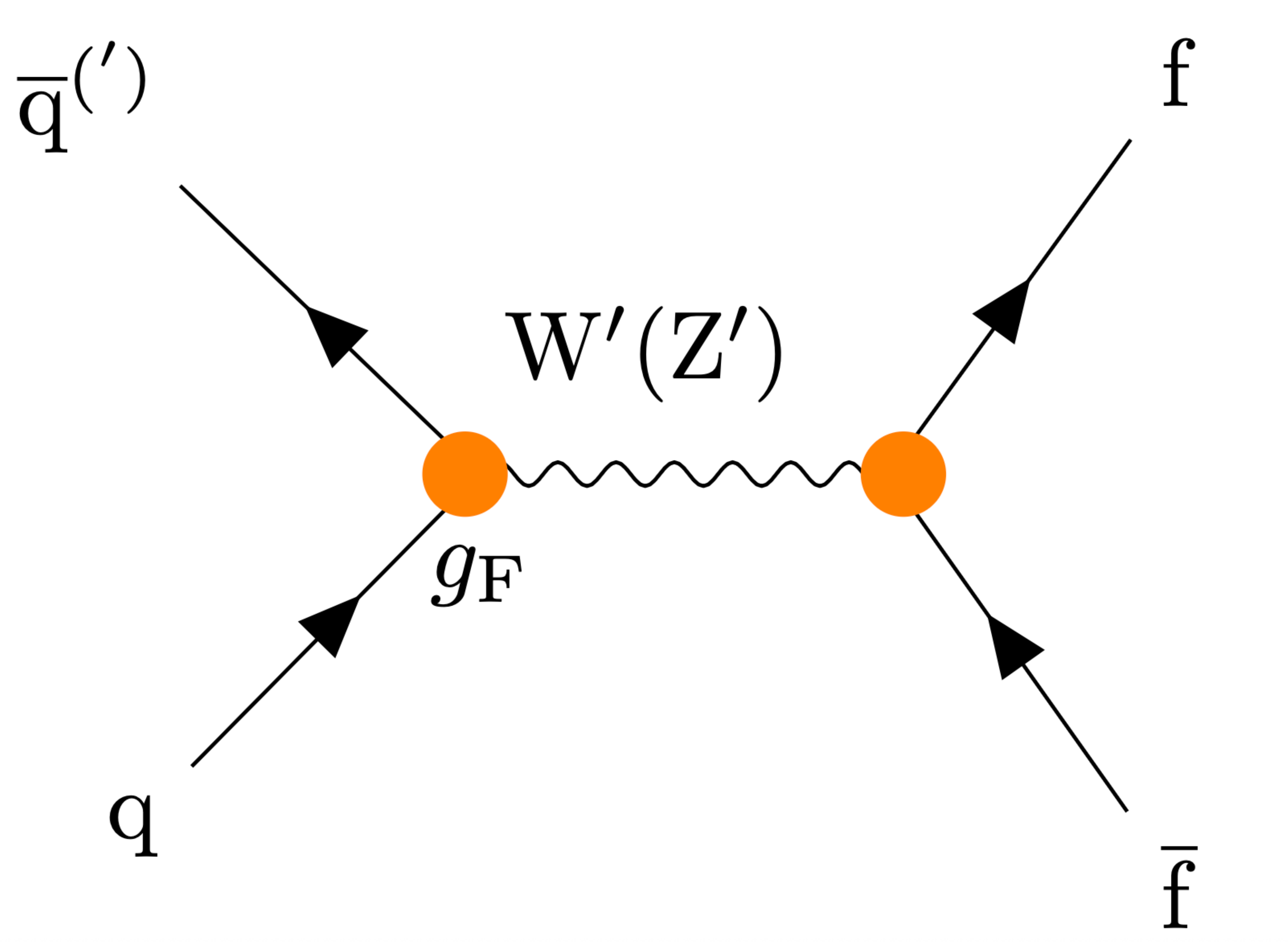} \\
\includegraphics[width=0.3\textwidth]{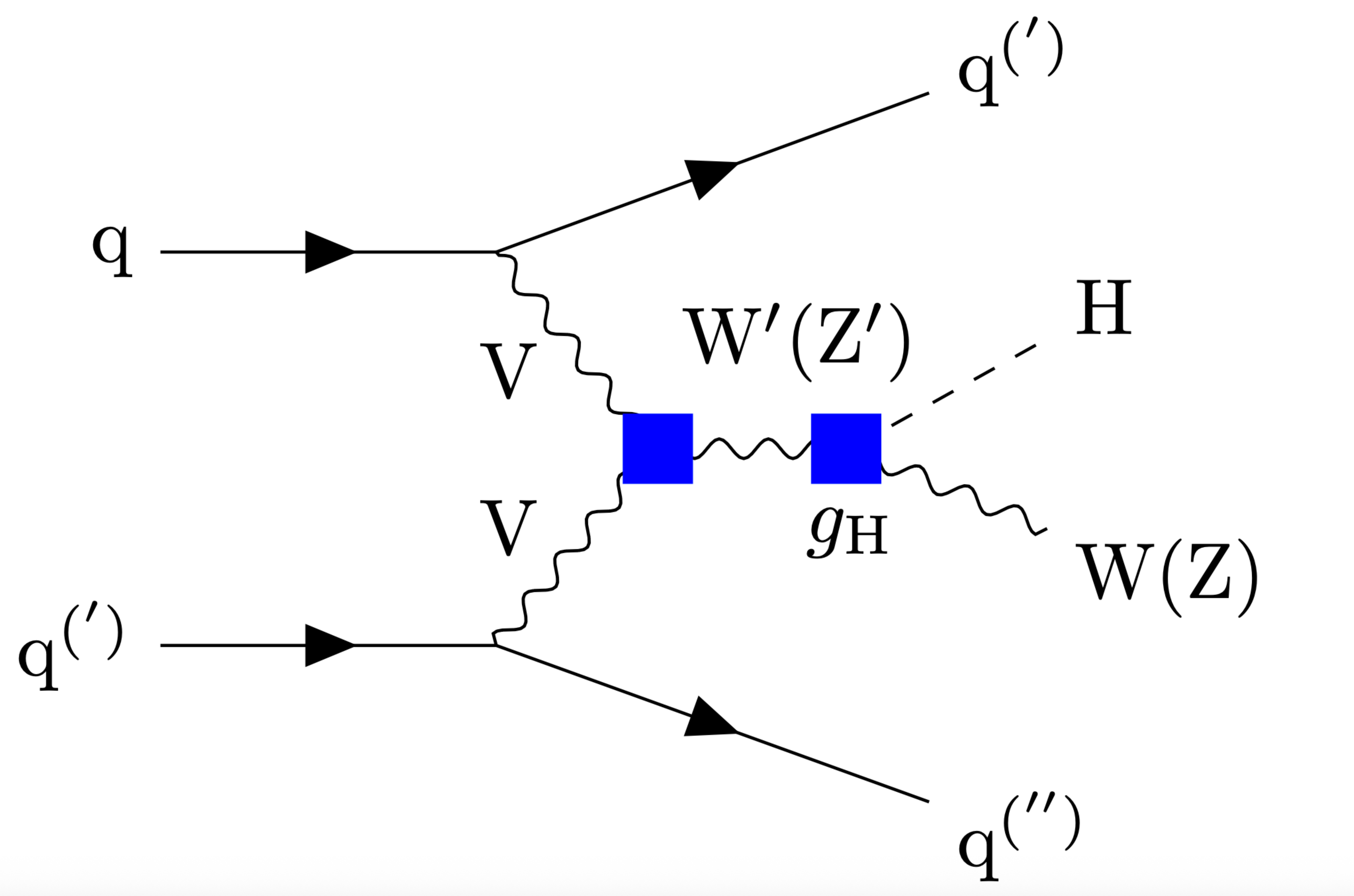}
\hspace{0.1\textwidth}
\includegraphics[width=0.3\textwidth]{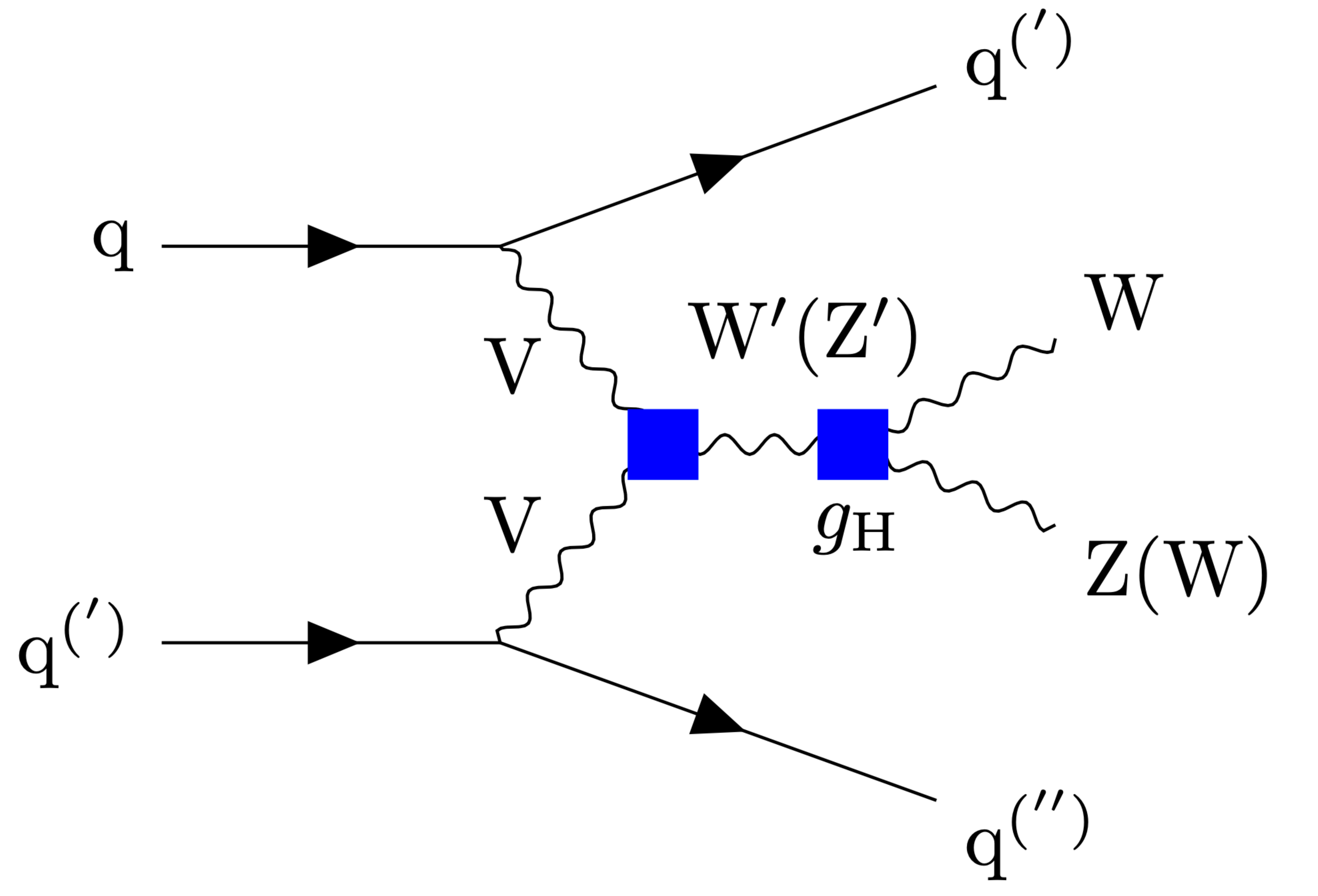}
\caption{
Representative Feynman diagrams for the production and decay of \Wprime and \Zprime bosons via the Drell--Yan process (upper row)
and the VBF process (lower row). The \Wprime (\Zprime) boson decays to \WH (\ZH) (upper left, lower left),
\WZ (\WW) (upper middle, lower right), and fermion-antifermion pairs (upper right).
The orange circle and the blue square indicate the fermionic and bosonic interaction vertices, respectively.
The definition of the couplings is provided in Section~\ref{sec:signals}.
}
\label{fig:diagrams}
\end{figure}

Depending on the underlying model, these heavy resonances could couple predominantly to SM fermions,
as predicted by minimal \Wprime and \Zprime models~\cite{Salvioni:2009mt,Schmaltz_2011}, or to SM bosons, as proposed in scenarios such as composite Higgs or little Higgs models~\cite{Schmaltz:2005ky,Perelstein:2005ka,Contino:2011np}.
In some cases, these resonances could act as a bridge, or portal, between SM and dark-matter particles,
providing experimental access to otherwise undetectable particles in the dark sector~\cite{Abercrombie:2015wmb}.
The heavy vector triplet (HVT) framework~\cite{Pappadopulo_2014} provides a
general parameterization for the production and decay of heavy spin-1 resonances that couple
to quarks, leptons, vector bosons, and \PH bosons.
Depending on the specific choice of the model parameters, the dominant production mechanism for a \Vprime boson at the LHC can be
either quark-antiquark annihilation through the Drell--Yan (DY) process, or vector boson fusion (VBF).
Representative Feynman diagrams for both production mechanisms and subsequent HVT boson decays are shown in Fig.~\ref{fig:diagrams}.

This paper presents a combined statistical analysis of CMS searches for heavy spin-1 resonances
decaying into pairs of bosons or fermions, interpreted in the context of the HVT framework.
A combined analysis of multiple channels is essential to maximize the sensitivity and fully exploit the outcome of these searches, particularly in light of localized excesses of events observed in individual channels.
For instance, a combined search by the CMS Collaboration for heavy resonance decays to \WW, \WZ, \ZZ, \WH, or \ZH in the fully hadronic final states~\cite{CMS:B2G-20-009},
based on a data sample of proton-proton (\pp) collisions at a center-of-mass energy of 13\TeV, corresponding to an integrated luminosity of 138\fbinv,
reported two excesses of events, each with a local significance of 3.6 standard deviations (\SD), at 2.1 and 2.9\TeV.
Similarly, in a combination of spin-1 resonance analyses using \pp collision data at the same center-of-mass energy, corresponding to an integrated luminosity of 139\fbinv,
the ATLAS Collaboration observed a local excess over the background prediction of 2.6 \SD at 1.5\TeV~\cite{ATLAS:2024qvg}.
While these deviations are not statistically significant enough to claim evidence for BSM physics, a combined interpretation including different analyses and final states
often improves the sensitivity of the search, providing a more robust assessment of potential deviations from the SM expectations.

The results presented here improve upon the previous CMS combination based on the 2016 data set~\cite{CMS:2019kaf}
and provide the most stringent constraints to date on BSM scenarios involving a triplet of heavy vector boson resonances.
These results represent a significant improvement over previous constraints from similar statistical combinations,
driven by the inclusion of a larger data set and advances in the underlying analyses.
All input analyses are based on the 2016--2018 data set, corresponding to an integrated luminosity of 138\fbinv
of \pp collisions at a center-of-mass energy of 13\TeV.
The combination includes the most sensitive channels for spin-1 resonance decays,
including \WW, \WZ, \WH, \ZH, and various fermion pairs such as \qqbar, \bbbar, \ttbar, \tb, \ellell, and \lnu, where $\Pell = \Pe, \PGm, \PGt$.
Charge-conjugate states are implied throughout the paper.
The tabulated results are provided in a HEPData record~\cite{HEPData}.

\section{The CMS detector and event reconstruction} \label{sec:CMS}

The central feature of the CMS apparatus is a superconducting solenoid of 6\unit{m} internal diameter, 
providing a magnetic field of 3.8\unit{T}. Within the solenoid volume are a silicon pixel and strip tracker, 
a lead tungstate crystal electromagnetic calorimeter, and a brass and scintillator hadron 
calorimeter, each composed of a barrel and two endcap sections. Forward calorimeters extend 
the pseudorapidity coverage provided by the barrel and endcap detectors. Muons are measured 
in gas-ionization detectors embedded in the steel flux-return yoke outside the solenoid.
More detailed descriptions of the CMS detector are provided in Refs.~\cite{CMS:2008xjf,CMS:2023gfb}.

Events of interest are selected using a two-tiered trigger system~\cite{CMS:2016ngn}. The first level (L1), composed of custom 
hardware processors, uses information from the calorimeters and muon detectors to select events at a rate 
of approximately 100\unit{kHz} with a fixed latency of about 4\unit{$\mu$s}~\cite{CMS:2020cmk}. The second level, known 
as the high-level trigger, consists of a farm of processors running a version of the full event 
reconstruction software optimized for fast processing, and reduces the event rate to a few \unit{kHz} 
before data storage~\cite{CMS:2016ngn,CMS:2024aqx}.

A particle-flow (PF) algorithm~\cite{CMS:2017yfk} aims to reconstruct and identify each individual particle (photon, electron, muon, and charged and neutral hadron) in an event,
with an optimized combination of information from the various elements of the CMS detector.
For each event, hadronic jets are clustered from these reconstructed particles using the infrared and collinear safe anti-\kt algorithm~\cite{Cacciari:2008gp, Cacciari:2011ma}.
Small- and large-radius jets are clustered with distance parameters of $R=0.4$ and $R=0.8$, respectively.
Jet momentum is determined as the vectorial sum of all particle momenta in the jet,
and is found from simulation to be, on average, within 5 to 10\% of the true momentum
over the entire \pt spectrum and detector acceptance. 

Additional \pp interactions within the same or nearby bunch crossings (pileup) can contribute additional tracks and calorimetric energy depositions, increasing the jet momentum.
Pileup effects are mitigated using dedicated algorithms such as the charged-hadron subtraction and the
pileup-per-particle identification (PUPPI) methods~\cite{Bertolini:2014bba,Sirunyan:2020foa}.
The PUPPI algorithm is applied to reduce the pileup dependence of the \ptvecmiss observable, 
by weighting particle momenta according to their probability of originating from the primary interaction
vertex~\cite{CMS:2019ctu}.

Jet energy corrections are derived from simulation studies so that the average measured energy of jets becomes identical
to that of jets at the particle level. In situ measurements of the momentum balance in dijet,
\Gjets, \Zjets, and multijet events are used to determine 
any residual differences between the jet energy scale in data and simulation,
and appropriate corrections are made~\cite{CMS:2016lmd}. Additional selection criteria
are applied to each jet to remove those originating from instrumental effects or
reconstruction failures~\cite{Sirunyan:2020foa}.

For small-radius jets, neural-network-based taggers calibrated using collision data are used to identify jets originating from the hadronization of \PQb quarks (\PQb tagging)~\cite{CMS:2017wtu,Bols:2020bkb,CMS-BTV-22-001}.
For large-radius jets, the soft-drop algorithm~\cite{Larkoski:2014wba} is used to identify the subjets of a high-\pt \PW/\PZ/\PH boson candidate jet.
This algorithm is applied to large-radius jets reclustered with the Cambridge--Aachen algorithm~\cite{Dokshitzer:1997in,Wobisch:1998wt},
using an angular exponent $\beta = 0$ and a soft-cutoff threshold $z_{\text{cut}}<0.1$,
which removes soft, wide-angle radiation from the jet.
Neural-network-based taggers calibrated using collision data~\cite{CMS:2020poo,Qu:2019gqs,CMS:2017wtu,CMS-BTV-22-001}
are used to assign each large-radius jet to one of the following classes: a single quark or gluon, a \PW or \PZ boson decaying to \qqbar,
an \PH or \PZ boson decaying to \bbbar or \ccbar, and a top quark decaying hadronically.

The missing transverse momentum vector \ptvecmiss is computed as the negative vector sum of
the PUPPI-weighted transverse momenta of all the PF candidates in an event;
its magnitude is denoted as \ptmiss~\cite{CMS:2019ctu}.
The \ptvecmiss is modified to account for corrections to the energy scale of the reconstructed jets in the event.
Anomalous high-\ptmiss events can be due to a variety of reconstruction failures, detector malfunctions or noncollision backgrounds.
Such events are rejected by event filters that are designed to identify more than 85--90\% of the spurious high-\ptmiss events with a misidentification rate less than 0.1\%~\cite{CMS:2019ctu}.

\section{Signal models and theoretical framework} \label{sec:signals}

Several theoretical models, including composite Higgs scenarios~\cite{Contino:2011np,Accomando:2012yg,Bellazzini:2014yua}
and sequential extensions of the SM~\cite{Barger:1980ix,Grojean:2011vu}, predict the existence of additional resonances with masses in the \TeV range.
The HVT framework~\cite{Pappadopulo_2014} is based on an effective Lagrangian that provides
a phenomenological description encompassing a broad class of these models.

The HVT framework introduces three new heavy bosons, \Wprimepm and \Zprime,
which are nearly degenerate in mass and have other properties analogous to those of the SM weak vector bosons.
It allows for the exploration of a broad range of couplings between these heavy bosons and SM fermions and bosons.
The relevant interaction Lagrangian is:
\begin{equation}
\mathcal{L}^{\text{int}}_{\mathcal{W}} = -\gq\,\mathcal{W}^a_{\mu}\overline{q}_{k}\gamma^{\mu}\frac{\sigma_a}{2}q_{k} - \gl\,\mathcal{W}^a_{\mu}\overline{l}_{k}\gamma^\mu\frac{\sigma_a}{2}l_k - \gH\,(\mathcal{W}^a_{\mu}\PGF^{\dagger}\frac{\sigma_a}{2}iD^{\mu}\PGF + \text{h.c.}),
\label{HVT_L}
\end{equation}
where $q_k$ and $l_k$ denote the left-handed quark and lepton field doublets of generation $k$ ($k = 1, 2, 3$);
\PGF is the Higgs doublet; $\sigma_a$ ($a = 1, 2, 3$) are the Pauli matrices; and \gq, \gl, and \gH represent the couplings of the heavy triplet field $\mathcal{W}$ to SM quarks, leptons, and the Higgs boson, respectively.
Right-handed fermions do not couple directly to $\mathcal{W}$.
The fermion couplings \gq and \gl are assumed to be generation-universal, unless stated otherwise.
The triplet also couples to the Higgs boson and, by the equivalence theorem, to longitudinally polarized \PW and \PZ bosons.

The interaction structure is defined by three main parameters: \gV, which sets the overall strength of the new interaction, and the dimensionless parameters \cH and \cF. The coupling to the Higgs and longitudinal vector bosons is given by $\gH = \gV \cH$, while the generation-universal coupling to fermions is given by $\gF = \gq = \gl = g^2 \cF / \gV$.
Here, $g = 2 \mW / v \approx0.65$, where \mW is the \PW boson mass and $v$ is the vacuum expectation value of the Higgs field~\cite{PDG2024}.

Within this framework, three benchmark models are defined to represent different production and decay scenarios~\cite{Pappadopulo_2014}, referred to as models~A, B, and C.
Models~A and B assume quark-antiquark annihilation via the DY process as the dominant production mechanism, while model~C assumes production exclusively through VBF.
The distinction between models~A and B lies in the relative strengths of the couplings to bosons and fermions.
Model~A represents a weakly coupled extension of the SM based on an extended gauge symmetry~\cite{Barger:1980ix}.
In model~A, the parameter values are set to $\gH = -0.56$, $\gF = -0.55$, and $\gV = 1$.
Model~B corresponds to a strongly coupled scenario, as in composite Higgs models~\cite{Contino:2011np}, with $\gH = -2.93$, $\gF = 0.15$, and $\gV = 3$.
Both model~A and~B parameter values are set to the explicit theory scenario described in Ref.~\cite{Pappadopulo_2014}.
In model~B, the \Wprime and \Zprime bosons have larger intrinsic widths than in model~A; however, they remain narrower than the experimental resolution, except in purely leptonic final states.
Over the explored parameter space, the relative width $\Gamma/m$ of the \Vprime resonances remains below 5\%.
For decays to fermion pairs, model~A yields larger branching fractions, with each lepton generation contributing about 4\%, compared to approximately 0.2\% in model~B.
Conversely, decays to \VV or \VH final states have branching fractions of about 2\% in model~A and nearly 50\% in model~B.
Lastly, model~C sets $\gH = 1,~3,\text{or}\, 6$ and $\gV = 1$, and assumes exclusive VBF production by setting $\gF = 0$, resulting in a \Vprime boson that does not couple to SM fermions.

The production cross sections for the DY and VBF processes scale with the squares of the relevant couplings:
\begin{equation}
\sigma_{\text{DY}} = \gF^2\, \sigmahat_{\text{DY}}, \quad \sigma_{\text{VBF}} = \gH^2\, \sigmahat_{\text{VBF}},
\end{equation}
where $\sigmahat_{\text{DY}}$ and $\sigmahat_{\text{VBF}}$ are the reduced cross sections calculated with the coupling parameters factored out.

\section{Input channels} \label{sec:channels}

The individual channels included in this statistical combination are briefly summarized in the following.
For comprehensive descriptions, the reader is referred to the original publications listed in Table~\ref{tab:analyses}.

\begin{table}[t]
\centering
\topcaption{Summary of the searches for difermion and diboson resonances used as inputs to this combined statistical analysis. 
The targeted production modes, resonance mass ranges, reconstructed particles, and references are listed.
In the ``Jets'' column, only jets from the heavy resonance decay are indicated.
Detailed definitions of reconstructed particles in the ``Leptons'' and ``Jets'' columns, such as \tauh and various types of jets, are provided later in the main text.}
\label{tab:analyses}
\cmsTable{
\begin{tabular}{lccccccc}  
Channel                 & Prod.\ modes & \mV range [\TeVns{}] & Leptons  & Jets & Reference \\
\hline
\noalign{\vskip 1pt}
\Zprimetautau   & DY        & 0.5--4.0 & $\Pell\tauh$ or 2\tauh & \NA                   & \cite{CMS:EXO-21-016} \\
\Zprimell       & DY        & 0.8--5.5 & 2\Pell            & \NA                   & \cite{CMS:EXO-19-019} \\
\Wprimelnu      & DY        & 0.8--6.0 & 1\Pell            & \NA                   & \cite{CMS:EXO-19-017} \\
\Wprimetaunu    & DY        & 0.8--6.0 & 1\tauh           & \NA                   & \cite{CMS:EXO-21-009} \\[\cmsTabSkip]
\Vprimeqq       & DY        & 1.8--8.7 & 0\Pell            & \jet\jet              & \cite{CMS:EXO-19-012} \\
\Wprimetb(0\Pell)       & DY        & 1.0--4.0 & 0\Pell            & \jettop\jetb          & \cite{CMS:B2G-20-005} \\
\Wprimetb(1\Pell)       & DY        & 2.0--6.0 & 1\Pell            & \jetb\jetb            & \cite{CMS:B2G-20-012} \\
\Zprimett(1\Pell)       & DY        & 0.8--5.0 & 1\Pell            & \jettop or \jet\jet\jetb & \cite{CMS:B2G-25-009}\\[\cmsTabSkip]
\ZprimeZH($2\Pell2\PQb$)      & DY + VBF  & 0.8--5.0 & 2\Pell            & \jetbb                & \cite{CMS:B2G-19-006} \\
\ZprimeZH($2\PGn2\PQb$)       & DY + VBF  & 0.8--5.0 & 0\Pell            & \jetbb                & \cite{CMS:B2G-19-006} \\
\ZprimeZH($2\Pell2\PQq$)      & DY        & 1.2--6.0 & 2\Pell            & \jetH                 & \cite{CMS:B2G-23-008} \\
\ZprimeZH($2\PGn2\PQq$)       & DY        & 1.2--6.0 & 0\Pell            & \jetH                 & \cite{CMS:B2G-23-008} \\
\WprimeWZ($2\PQq2\Pell$)      & DY        & 0.6--2.0 & 2\Pell            & \jetV or \jet\jet        & \cite{CMS:B2G-20-013} \\
\WprimeWZ($2\PQq2\PGn$)       & DY + VBF  & 1.0--4.5 & 0\Pell            & \jetV                 & \cite{CMS:B2G-20-008} \\
\VprimeWVWH     & DY + VBF  & 1.0--4.5 & 1\Pell            & \jetV                 & \cite{CMS:B2G-19-002} \\
\VprimeVVVH     & DY + VBF  & 1.3--6.0 & 0\Pell            & \jetV\jetV            & \cite{CMS:B2G-20-009} \\
\end{tabular}
}
\end{table}

This combined statistical analysis targets heavy resonances with masses between 1 and 8\TeV.
In this regime, the \PW, \PZ, and \PH bosons, as well as top quarks produced in \Vprime boson decays, typically have a high Lorentz boost in the laboratory frame.
When these bosons decay hadronically, their decay products are often collimated and reconstructed as single large-radius jets, which are identified using jet substructure and machine-learning-based algorithms.
In most searches, the resonance mass is reconstructed from a pair of such large-radius jets.
Exceptions include channels, such as \WprimeWZ~\cite{CMS:B2G-20-013} and \Zprimett(1\Pell)~\cite{CMS:B2G-25-009},
where hadronic reconstructed candidates are also resolved as multiple small-radius jets,
and channels \Wprimelnu~\cite{CMS:EXO-19-017}, \Wprimetaunu~\cite{CMS:EXO-21-009}, \ZprimeZH~\cite{CMS:B2G-19-006,CMS:B2G-23-008}, and \WprimeWZ~\cite{CMS:B2G-20-008},
where the presence of neutrinos from the \PZ boson decay prevents full mass reconstruction, and the transverse mass is used instead.
Several channels of diboson final states further enhance sensitivity to the VBF production mode by selecting events with two additional small-radius jets that have a large invariant mass and a large separation in pseudorapidity.

The analysis channels are categorized according to the reconstructed final-state candidates:

\begin{itemize}
\item \Pell: muons or electrons,
\item \tauh: hadronically decaying \PGt leptons,
\item \jet: small-radius jets,
\item \jetb: small-radius jets identified as originating from \PQb quarks (\btagged),
\item \jetV: large-radius jets   identified as \PW or \PZ bosons decaying to \qqbar (\Vtagged),
\item \jetbb: large-radius jets  identified as \PH or \PZ bosons decaying to \bbbar (\bbtagged),
\item \jetH: large-radius jets   identified as \PH bosons decaying hadronically (\Htagged), and 
\item \jettop: large-radius jets identified as hadronically decaying top quarks (\ttagged).
\end{itemize}

In all channels, the signal is searched for as a localized excess over a smoothly falling background in the reconstructed mass or transverse mass spectrum.
Background estimation strategies vary by channel.
Analyses involving leptons and small-radius jets primarily rely on Monte Carlo (MC) simulations to model the background.
In contrast, analyses that use neural-network-based, large-radius jet taggers employ control samples in data to estimate the background, as the misidentification rates for these taggers are challenging to simulate accurately.

To ensure statistical independence among channels, event selections were designed to avoid overlaps in both the signal and control regions.
Mutual exclusivity is enforced in most channels through requirements on lepton and jet multiplicities.
If overlaps are identified, the overlapping phase space is retained in the combination only in the channel that provides the highest expected sensitivity for a given signal model and mass hypothesis.
This procedure is implemented such that signal regions from analyses that are less sensitive in the overlapping phase space are not included in the statistical combination for that specific signal model and mass point.

In channels targeting fermionic decays of \Vprime bosons, overlaps primarily occur among fully hadronic final states.
The inclusive dijet search \Vprimeqq~\cite{CMS:EXO-19-012}, which requires only two high-\pt jets, overlaps with all other fully hadronic channels because of its inclusive event selection and is therefore excluded whenever a more sensitive channel covers the same phase space.
Two additional searches for \Zprimebb~\cite{CMS:EXO-20-008} and \Zprimett(0\Pell)~\cite{CMS:B2G-25-009} are not included in this combination because they have a large overlap with the inclusive \Vprimeqq search~\cite{CMS:EXO-19-012}.
These channels are specifically optimized to study exclusive final states. 
They provide enhanced sensitivity relative to the inclusive \Vprimeqq search for BSM scenarios other than the ones probed in this combination.
Overlaps also occur in channels with one lepton, such as \Wprimetb~\cite{CMS:B2G-20-012} and \Zprimett~\cite{CMS:B2G-25-009}, where one of the top quarks decays leptonically.

In the diboson channels, each channel requires either at least two high-\pt small-radius jets or a single large-radius jet consistent with a hadronically decaying boson.
Event categories are constructed to be mutually exclusive using selections based on lepton and jet multiplicities, as well as the invariant mass and jet tagging requirements of final-state candidates.
In the \ZprimeZH($2\Pell2\PQb+2\PGn2\PQb$)~\cite{CMS:B2G-19-006} channel, events are categorized based on the number of \btagged subjets within the large-radius jet, defining a 2-\PQb-tag and a ${\leq}1$-\PQb-tag categories.
The latter category overlaps with the signal region of the \ZprimeZH($2\Pell2\PQq+2\PGn2\PQq$)~\cite{CMS:B2G-23-008} channel and is not included in the combination, as the signal regions of Ref.~\cite{CMS:B2G-23-008} provide greater sensitivity in this phase space. 
In contrast, events with two \btagged jets of Ref.~\cite{CMS:B2G-23-008} are vetoed, ensuring that the selected events are statistically independent of the 2-\PQb-tag category of Ref.~\cite{CMS:B2G-19-006}. Thus, only nonoverlapping signal regions are retained in the statistical combination.
In addition, the \WprimeWZ~\cite{CMS:B2G-20-013} channel uses a 2-\PQb-tag category as a control region, which overlaps with the signal region of the \ZprimeZH($2\Pell2\PQb+2\PGn2\PQb$)~\cite{CMS:B2G-19-006} channel.
To maintain statistical independence, the \btagged categories of the \WprimeWZ~\cite{CMS:B2G-20-013} channel are therefore excluded from the combination.

\section{Systematic uncertainties}

An overview of the main experimental and theoretical systematic uncertainties considered in the analyses included in the combination
is provided in the following.
Detailed descriptions can be found in the original publications listed in Table~\ref{tab:analyses}.
The sensitivity of the combined analysis is limited by the statistical uncertainty,
with all systematic uncertainties being subdominant.
A summary of the main systematic uncertainties is presented in Table~\ref{tab:systematics}.

\begin{table}[t]
\centering
\topcaption{Summary of the main systematic uncertainties.
The second column specifies whether the uncertainty affects the yield (Y), the shape (\SP) of the distributions, or both.
The ``Effects'' columns summarize the typical impact of the uncertainty on the yield, migration, and signal shape parameters.
Analyses are split by fermion and boson signatures.
The letters ``\SI'', ``\BA'', and ``\SB'' indicate whether the uncertainty affects the signal, the dominant background, or both, respectively.
The letter ``\UC'' denotes cases where background parameters are unconstrained and allowed to float freely in the fit.
A dash ``\NA'' refers to cases where an uncertainty has a negligible effect or is not applicable.}
\cmsTable{
\begin{tabular}{lc  ccccccccc ccccccc}
                                   & Type  & Effects & \rotatebox{90}{\Zprimetautau~\cite{CMS:EXO-21-016}} & \rotatebox{90}{\Zprimell~\cite{CMS:EXO-19-019}} & \rotatebox{90}{\Wprimelnu~\cite{CMS:EXO-19-017}} & \rotatebox{90}{\Wprimetaunu~\cite{CMS:EXO-21-009}} & \rotatebox{90}{\Vprimeqq~\cite{CMS:EXO-19-012}} & \rotatebox{90}{\Wprimetb~\cite{CMS:B2G-20-005}} & \rotatebox{90}{\Wprimetb~\cite{CMS:B2G-20-012}} & \rotatebox{90}{\Zprimett~\cite{CMS:B2G-25-009}} & Effects & \rotatebox{90}{\ZprimeZH~\cite{CMS:B2G-19-006}} & \rotatebox{90}{\ZprimeZH~\cite{CMS:B2G-23-008}} & \rotatebox{90}{\WprimeWZ~\cite{CMS:B2G-20-013}} & \rotatebox{90}{\WprimeWZ~\cite{CMS:B2G-20-008}} & \rotatebox{90}{\VprimeWVWH~\cite{CMS:B2G-19-002}} & \rotatebox{90}{\VprimeVVVH~\cite{CMS:B2G-20-009}} \\
\hline
\noalign{\vskip 1pt}
Sig. PDF and scale                 & Y, \SP  & 2--20\%     & \SI & \NA & \NA & \SI & \NA & \SI & \NA & \SI & 0.3--30\%  & \SI & \SI & \SI & \SI & \SI & \SI \\
Bkg. PDF and scale                 & Y, \SP  & 2--50\%     & \BA & \NA & \BA & \BA & \NA & \BA & \BA & \BA & 0.1--20\%  & \BA & \NA & \NA & \BA & \NA & \NA \\[\cmsTabSkip]
Bkg. modeling                      & Y, \SP  & 1--100\%    & \BA & \UC & \BA & \BA & \UC & \BA & \BA & \BA & 4--50\%    & \UC & \UC & \BA & \BA & \BA & \BA \\
Jet energy scale/res.              & Y, \SP  & 1--40\%     & \NA & \NA & \SB & \SB & \SI & \SB & \SB & \SB & 0.2--35\%  & \SB & \SI & \SB & \SB & \SI & \SB \\
Jet mass scale/res.                & Y, \SP  & 2--5\%      & \NA & \NA & \NA & \NA & \NA & \NA & \NA & \NA & 0.3--2.6\% & \SB & \NA & \SB & \SB & \SB & \SI \\
\PH tagging                        & Y     & \NA         & \NA & \NA & \NA & \NA & \NA & \NA & \NA & \NA & 0.4--15\%  & \SI & \SI & \NA & \NA & \SI & \SB \\
\PV tagging                        & Y     & \NA         & \NA & \NA & \NA & \NA & \NA & \NA & \NA & \NA & 1.2--33\%  & \NA & \NA & \SB & \SB & \SI & \SB \\
\PQt tagging                       & Y, \SP  & 21--36\%    & \NA & \NA & \NA & \NA & \NA & \NA & \NA & \SB & 1--3\%     & \NA & \NA & \NA & \NA & \NA & \BA \\
\PQb tagging and veto              & Y, \SP  & 1--23\%     & \SB & \NA & \NA & \NA & \NA & \SB & \SB & \SB & 0.1--15\%  & \SB & \SI & \SB & \SB & \SI & \NA \\[\cmsTabSkip]
\Pe, \PGm id., iso., trigger       & Y, \SP  & 0.5--8\%    & \SB & \SI & \SB & \SB & \NA & \NA & \SB & \SB & 0.9--7\%   & \SB & \SI & \SB & \NA & \SB & \NA \\
\Pe, \PGm energy scale/res.        & Y, \SP  & 1--6\%      & \SI & \SI & \SB & \SB & \NA & \NA & \SB & \NA & 0.1--3\%   & \SI & \NA & \SB & \NA & \SI & \NA \\
\tauh reco., id., iso.             & Y     & 5--10\%     & \SI & \NA & \NA & \SB & \NA & \NA & \NA & \NA & \NA        & \NA & \NA & \NA & \NA & \NA & \NA \\
\tauh energy scale/res.            & Y, \SP  & 1.5--5\%    & \SI & \NA & \NA & \SB & \NA & \NA & \NA & \NA & \NA        & \NA & \NA & \NA & \NA & \NA & \NA \\[\cmsTabSkip]
Int. luminosity                    & Y     & 1.2--2.5\%  & \SB & \SI & \SB & \SB & \SI & \SB & \SB & \SB & 1.2--2.5\% & \SB & \SI & \SB & \SB & \SI & \SB \\
\ptmiss scale	                   & Y, \SP  & 2--5\%      & \NA & \NA & \SB & \SB & \NA & \NA & \NA & \NA & 0.1--2.4\% & \SB & \NA & \NA & \SB & \SI & \NA \\
Pileup                             & Y, \SP  & 0.6--5\%    & \NA & \NA & \SB & \SB & \NA & \SB & \SB & \SB & 0.1--4\%   & \SB & \SI & \SB & \SB & \SI & \SB \\
L1 trigger inefficiency            & Y, \SP  & 0.5--5\%    & \NA & \NA & \SB & \SB & \NA & \SB & \NA & \NA & 0.2--1\%   & \NA & \SI & \SB & \SB & \NA & \SB \\
\end{tabular}
}
\label{tab:systematics}
\end{table}

Theoretical uncertainties arise from variations in the renormalization and factorization scales used in the matrix element calculations, as well as from the choice of the parton distribution functions (PDFs) and their internal parameters.
The PDF uncertainties are evaluated with either the NNPDF3.0~\cite{NNPDF:2014otw} or NNPDF3.1~\cite{NNPDF:2017mvq} sets, depending on the analysis.
Uncertainties from scale variations are treated as correlated across all analyses.
The PDF uncertainties are treated as correlated among analyses that use the same PDF set and as uncorrelated otherwise.
The impact of these theoretical uncertainties is separated into the acceptance and normalization effects.
In analyses where the acceptance uncertainty for the signal was found to be negligible, it is not included.
The normalization components of these uncertainties for the signal are included in the theoretical cross section predictions for each model, following the prescription of Ref.~\cite{Pappadopulo_2014}.

Most analyses rely on simulation to model background processes,
incorporating theoretical uncertainties from PDFs and from scale variations.
Minor backgrounds, such as single top quark and diboson processes, are typically subdominant and are assigned inclusive cross section uncertainties.
The treatment of subdominant backgrounds varies across analyses and is not detailed here.
The modeling of the QCD multijet background also differs between analyses.
For instance, in the \Vprimeqq analysis, this background is predicted from the data by fitting the dijet invariant mass distribution, while in the fully hadronic diboson channels, it is estimated using simulation.
Because of differences in phase space definitions and normalization strategies, background-related uncertainties are generally treated as uncorrelated across analyses, even when originating from the same underlying physics process.

Jet-related systematic uncertainties play a critical role in analyses
involving hadronic or lepton+jets final states, where precise jet reconstruction and flavor tagging are essential for signal
identification and background suppression.
For small-radius jets, the dominant sources of uncertainty include variations in the jet energy scale and resolution.
For large-radius jets, uncertainties in the jet energy and mass scales,
as well as their respective resolutions, are particularly important.
These uncertainties are propagated to both signal and background processes and are treated as correlated across analyses
that use the same jet type.
Jet-tagging uncertainties, including those for \PH, \PV, \PQb, and \PQt tagging, are derived from measurements
of tagging efficiencies and misidentification rates.
Uncertainties associated with the same tagging algorithm and calibration strategy are treated as correlated across analyses.

For electrons and muons, systematic uncertainties arise from
the energy or momentum scale and resolution measurements, as well as from the reconstruction, trigger, and identification efficiencies.
These uncertainties are typically correlated between signal and background, and also across the analyses
that use the same lepton selection criteria.
Lepton-related uncertainties are particularly important in the \Wprimelnu and \Zprimell analyses, as detector resolution and energy scale calibration significantly impact the modeling of the $\PW \to \lnu$ and $\PZ \to \ellell$ backgrounds.

Additional systematic uncertainties arise from the energy scale of unclustered particles contributing to \ptmiss,
the integrated luminosity, pileup modeling, and the inefficiency of the L1 trigger due to detector timing drift~\cite{CMS:2020cmk}.
The uncertainties in the integrated luminosity for the 2016, 2017, and 2018 data-taking periods range from 1.2 to 2.5\%~\cite{CMS:2021xjt,CMS-PAS-LUM-17-004,CMS-PAS-LUM-18-002}, with a combined uncertainty of 1.6\% for the full data set.

\section{Results and interpretation} \label{sec:Results}

The combined results are obtained through a simultaneous maximum likelihood fit across the individual analyses using the CMS statistical analysis tool \textsc{Combine}~\cite{CAT-23-001}, which is based on the \textsc{RooFit}~\cite{Verkerke:2003ir} and \textsc{RooStats}~\cite{Moneta:2010pm} frameworks.
In each individual channel, the data are found to be globally in agreement with the expectations from the SM.
A goodness-of-fit test of the combined analysis, based on a saturated test statistic~\cite{Cousins:2018tiz}, yields a $p$-value of approximately 34\% under the background-only hypothesis. This result indicates that the data are compatible with the SM prediction.

The observed and expected 95\% confidence level (\CL) upper limits on the signal production cross section and other parameters of interest (POIs) are computed using a modified frequentist \CLs method~\cite{CLS1,CLS2}, with an asymptotic approximation~\cite{Cowan:2010js}. This approach uses a profile likelihood ratio test statistic where systematic uncertainties are modeled as nuisance parameters~\cite{CAT-23-001}.
The POIs include the signal strength ($\mu$), defined as the ratio of the measured product of cross section and branching fraction ($\sigma\mathcal{B}$) to its theoretical expectation, and the coupling strength parameters (\gF, \gV, \gH).
Upper limits on the production cross section or couplings strengths are derived as functions of the resonance mass.
The evolution of the expected signal shape, acceptance, and branching fraction with the mass hypothesis is fully incorporated into the statistical model.

In certain channels, the asymptotic approximation becomes inaccurate in low-yield regimes, particularly in the high-mass tails of the reconstructed mass distributions.
In these cases, MC pseudo-experiments are performed to validate the accuracy of the asymptotic method, with nuisance parameters related to systematic uncertainties fixed to their post-fit values.
For the \Zprimell and \Wprimelnu channels in the high-mass regime (above 4\TeV), the asymptotic approximation is found to overestimate the sensitivity by approximately 15\% compared to the results obtained from MC pseudo-experiments.
This discrepancy is negligible in all other channels.

\begin{figure}[t]
\centering
\includegraphics[width=0.6\textwidth]{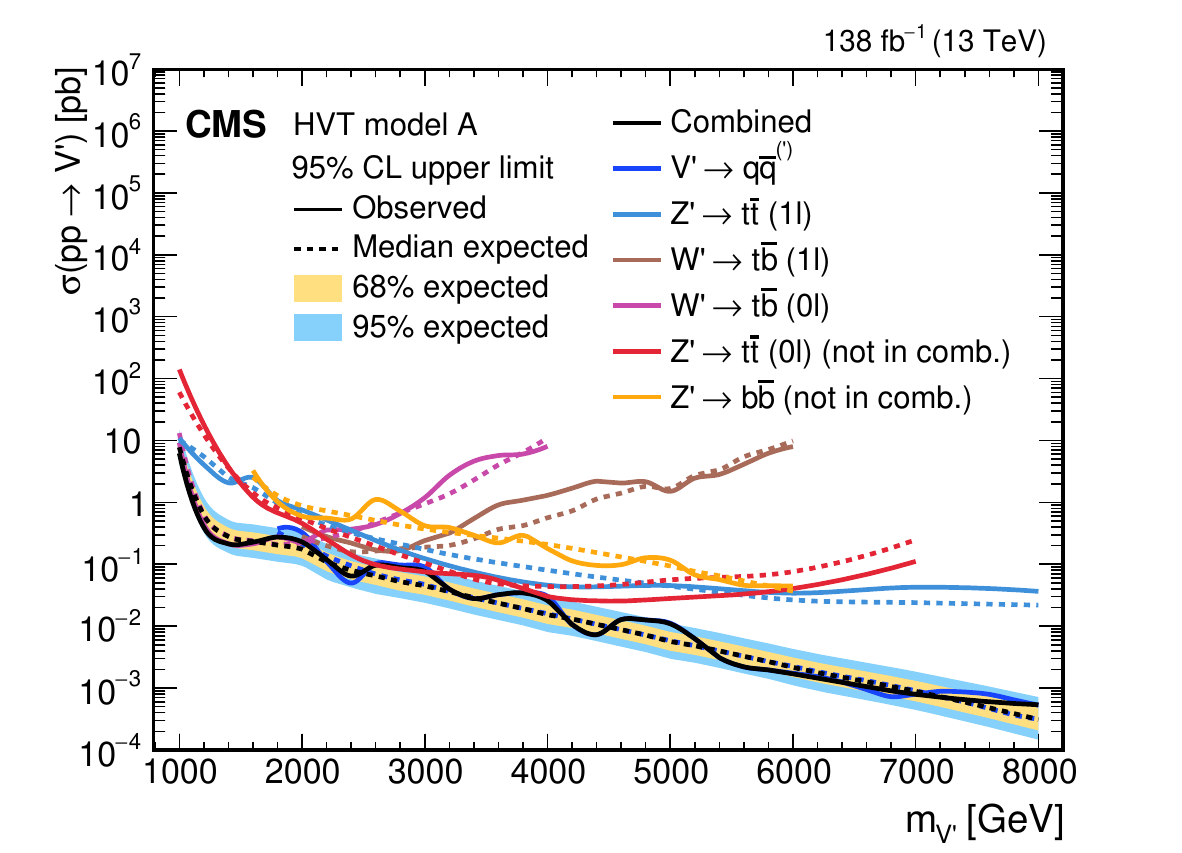} \\
\includegraphics[width=0.6\textwidth]{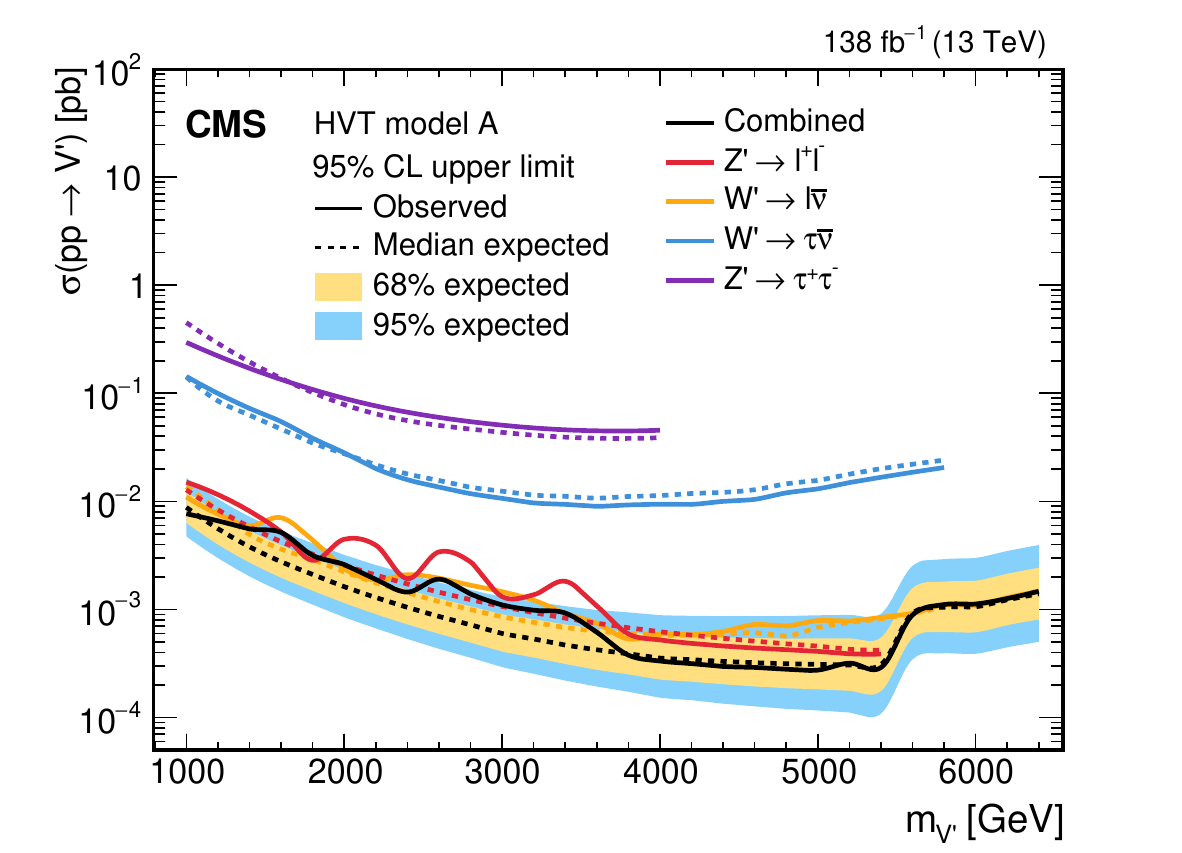}
\caption{Expected and observed 95\% \CL upper limits on the \Vprime boson production cross section as functions of the resonance mass \mV shown separately for the \Vprimequarks (upper) and \Vprimeleptons (lower) categories. The limits are evaluated in the HVT model~A scenario.
}
\label{fig:1d_individual_AB}
\end{figure}

\begin{figure}[th]
\centering
\includegraphics[width=0.6\textwidth]{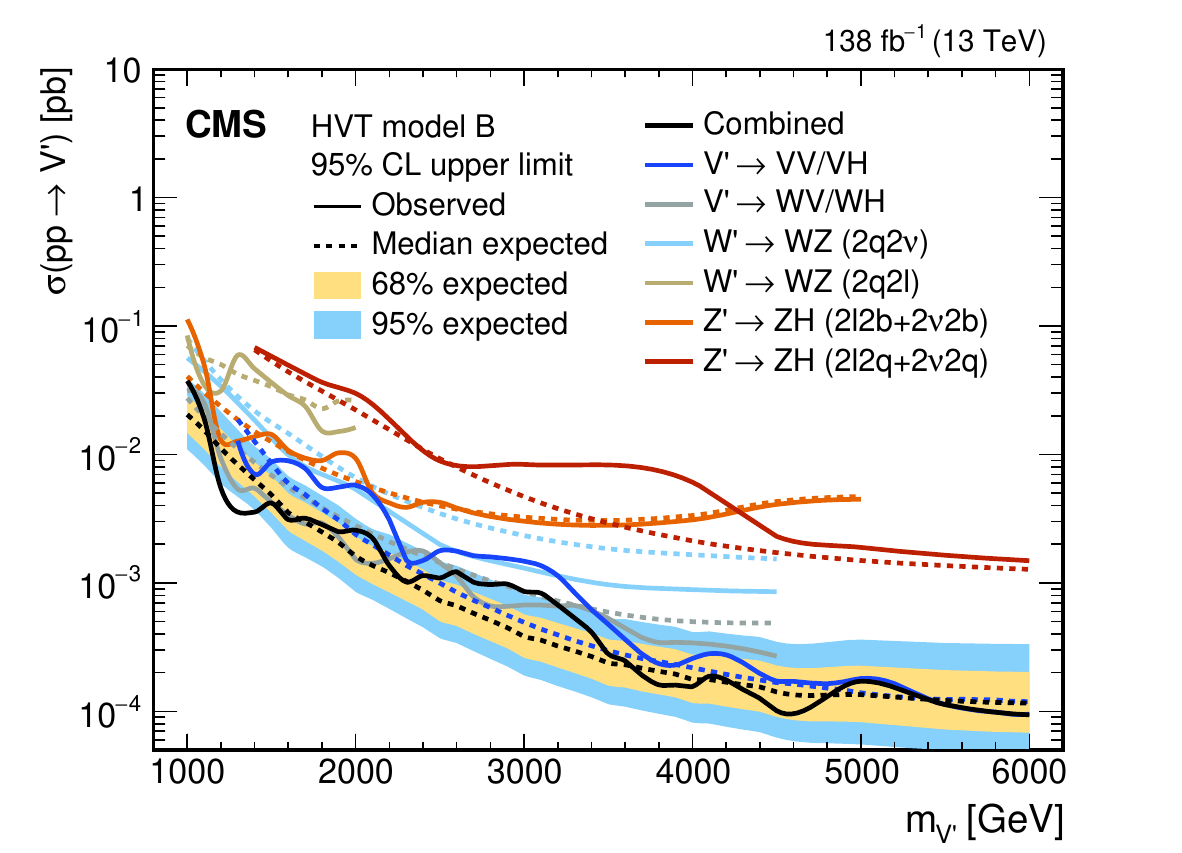}
\caption{Expected and observed 95\% \CL upper limits on the \Vprime boson production cross section as functions of the resonance mass \mV shown separately for the \Vprimebosons category. The limits are evaluated under the HVT model~B scenario.
}
\label{fig:1d_individual_C}
\end{figure}

Figures~\ref{fig:1d_individual_AB} and \ref{fig:1d_individual_C} present the expected and observed 95\% \CL upper limits on the \Vprime boson production cross section
as functions of the resonance mass \mV for the individual analyses targeting hadronic, leptonic, and bosonic final states, as well as for their combination.
The analyses are categorized according to the \Vprime boson decay mode, denoted as \Vprimequarks, \Vprimeleptons, and \Vprimebosons.
The limits for the \Vprimequarks and \Vprimeleptons categories are evaluated under the HVT model~A scenario, while the \Vprimebosons limits assume model~B.
The dominant sensitivity in the \Vprimequarks category arises from the fully hadronic \Wprimetb channel for resonance masses below 2\TeV and from the \Vprimeqq channel at higher masses.
In the \Vprimeleptons category, the leading contributions come from the \Wprimelnu and \Zprimell channels across the entire mass range.
For \Vprimebosons category, the sensitivity is primarily driven by the fully hadronic \VV and \VH channels.

\begin{figure}[th]
\centering
\includegraphics[width=0.6\textwidth]{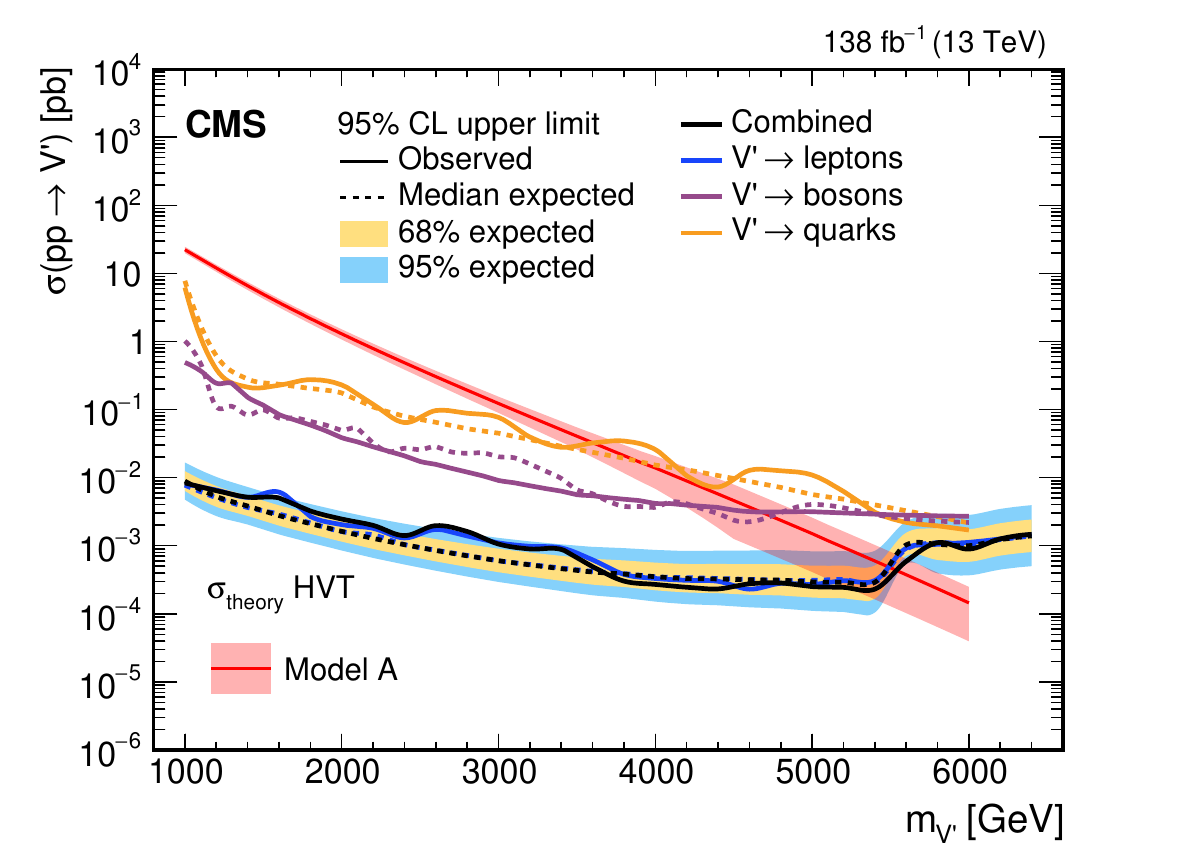} \\
\includegraphics[width=0.6\textwidth]{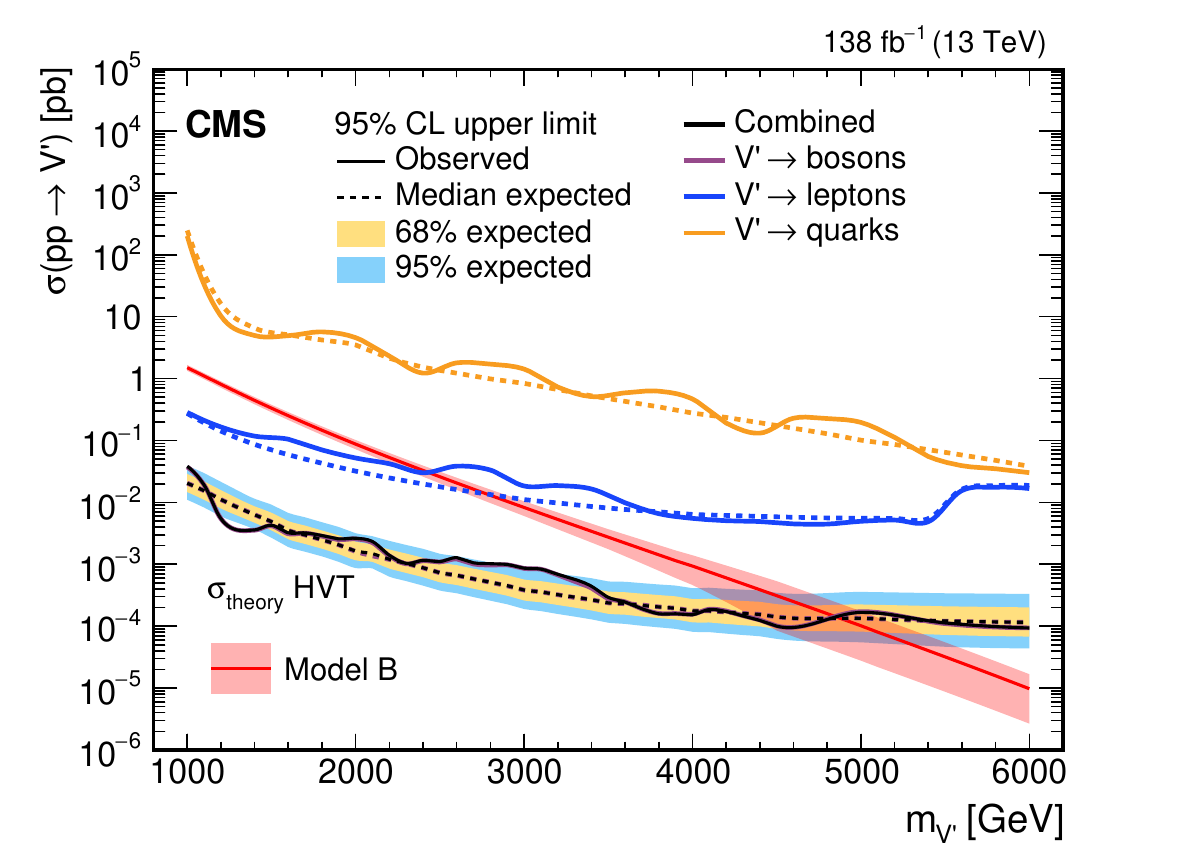}
\caption{Expected and observed 95\% \CL upper limits on the \Vprime boson production cross section
as functions of the resonance mass \mV in the HVT models~A (upper) and~B (lower).
The line corresponding to the \Vprimebosons category almost entirely overlaps with the combined result in the lower plot.}
\label{fig:1d_combined_AB}
\end{figure}

\begin{figure}[th]
\centering
\includegraphics[width=0.6\textwidth]{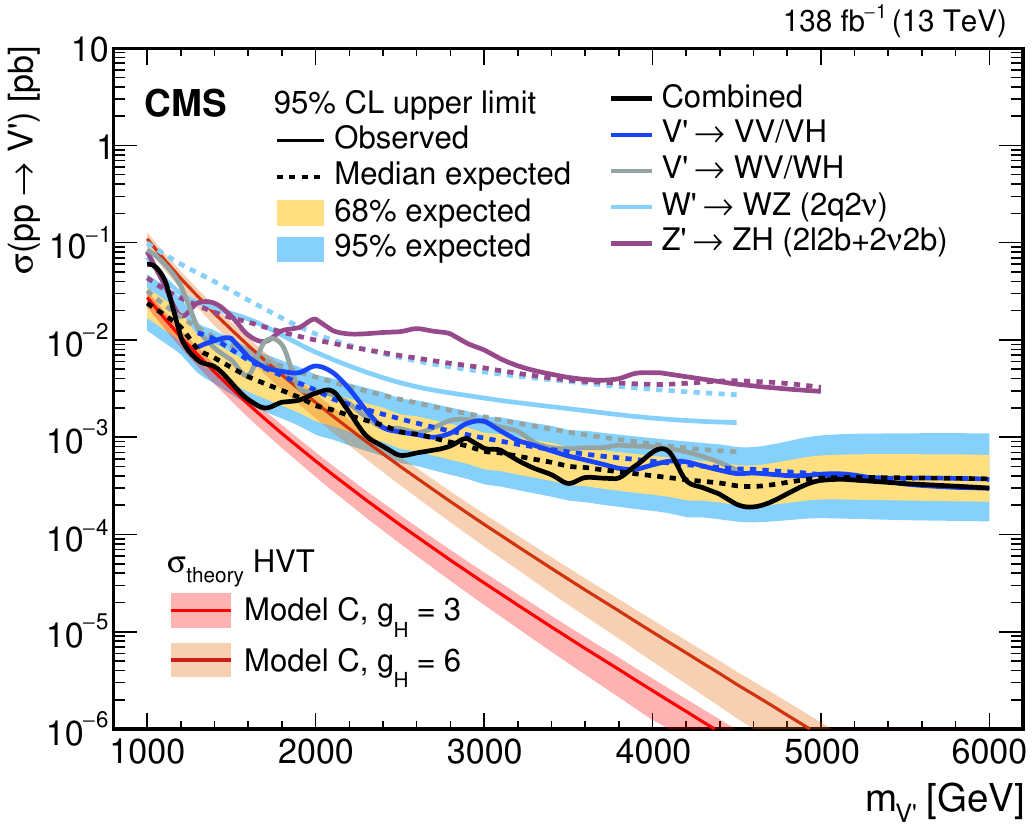}
\caption{Expected and observed 95\% \CL upper limits on the \Vprime boson production cross section
as functions of the resonance mass \mV in the HVT model~C.}
\label{fig:1d_combined_C}
\end{figure}

Figures~\ref{fig:1d_combined_AB} and \ref{fig:1d_combined_C} show the combined limits from all contributing channels for each HVT model.
For model~A, the dominant contribution from leptonic channels excludes $\Vprime$ resonances with masses below 5.5\TeV.
Above this mass, the sensitivity of the \Zprimell channel is statistically limited,
and the combined limit is driven by the \Wprimelnu channel.
For model~B, the bosonic channels dominate the sensitivity, excluding masses below 4.8\TeV.
In model~C, where only bosonic decays contribute, masses between 1.2 and 1.4\TeV are excluded for $\gH=3$, and masses below 2\TeV are excluded for $\gH=6$.
A decomposition of the combined limits into the individual \Wprime and \Zprime contributions is provided in Appendix~\ref{app:suppMat}.

No significant excess above the SM expectation is observed in the combined analysis.
The most significant local excesses in individual channels, those with a local significance greater than 2 \SD, were not consistent across channels, leading to a reduction of their significance in the combined result.
The largest excesses observed in model~A have local significances of 2.4 \SD and 2.1 \SD at masses of 2.6 and 3.4\TeV, respectively.
These excesses are primarily driven by the \Zprimell channel, and their significance is reduced in the combination compared to the previous CMS result~\cite{CMS:EXO-19-019}.
For model~B, the largest excesses have a local significance of 2.3 \SD at 2.9\TeV and 2.4 \SD at 3.0\TeV, dominated by the \Vprimebosons categories.
This combination also reduces the local significance of the previously reported CMS excesses of 3.6 \SD at 2.1 and 2.9\TeV~\cite{CMS:B2G-20-009}.
The largest excess in model~C, with a local significance of 3.0 \SD at a mass of 1\TeV in the \WV and \WH final states~\cite{CMS:B2G-19-002}, is reduced to 2.8 \SD in the combination.
The combination does not confirm the excess of events over the background prediction previously reported by the ATLAS Collaboration in the region around 1.5\TeV~\cite{ATLAS:2024qvg}.

Overall, this combination sets the most stringent limits to date for HVT models~B and C.
The improved sensitivity is primarily driven by the fully hadronic final states~\cite{CMS:B2G-20-009},
which benefit from the use of advanced jet substructure tagging techniques and a three-dimensional signal extraction strategy.
The combined limits in model~A, whose sensitivity is instead dominated by the leptonic channels, are comparable with those obtained by the ATLAS Collaboration~\cite{ATLAS:2024qvg}.

\begin{figure}[!ht]
\centering
\includegraphics[width=0.49\textwidth]{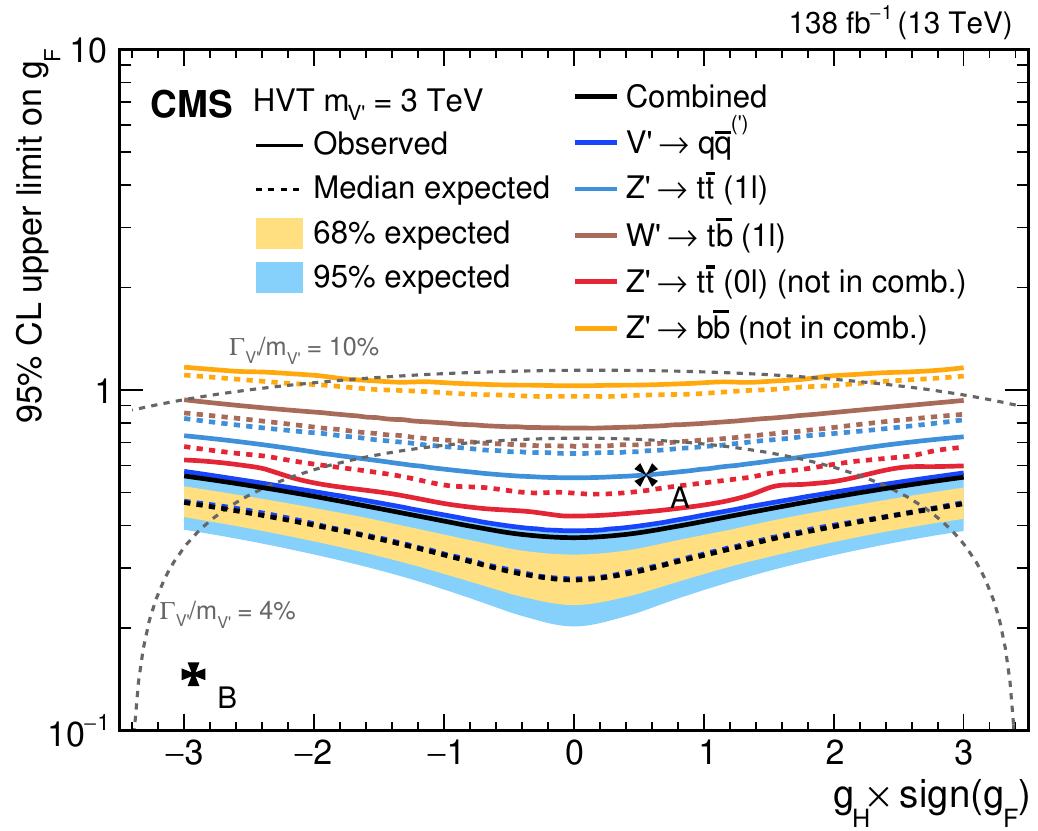}
\includegraphics[width=0.49\textwidth]{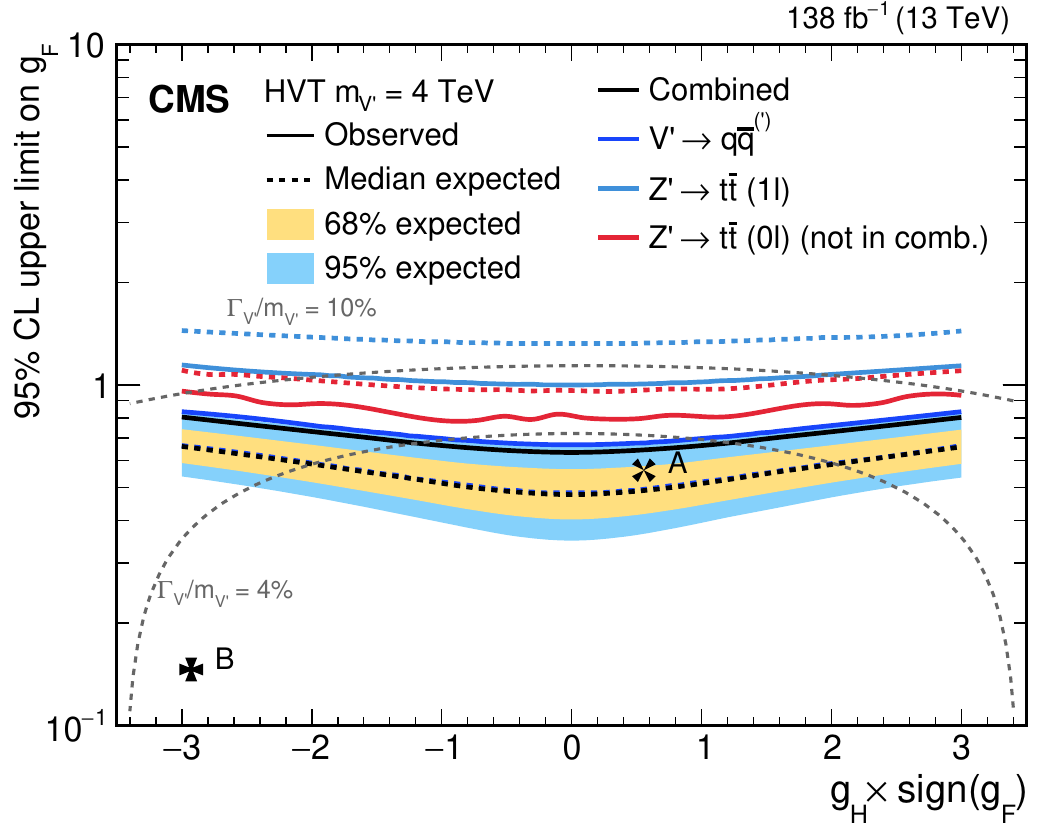}\\
\includegraphics[width=0.49\textwidth]{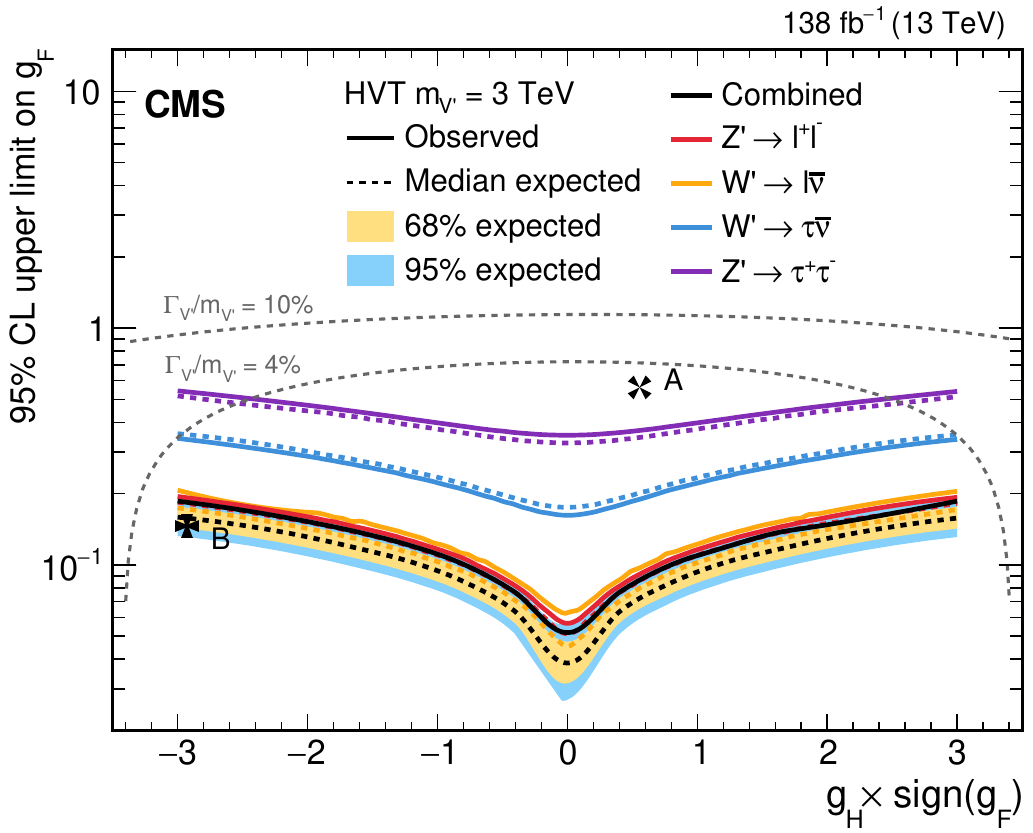}
\includegraphics[width=0.49\textwidth]{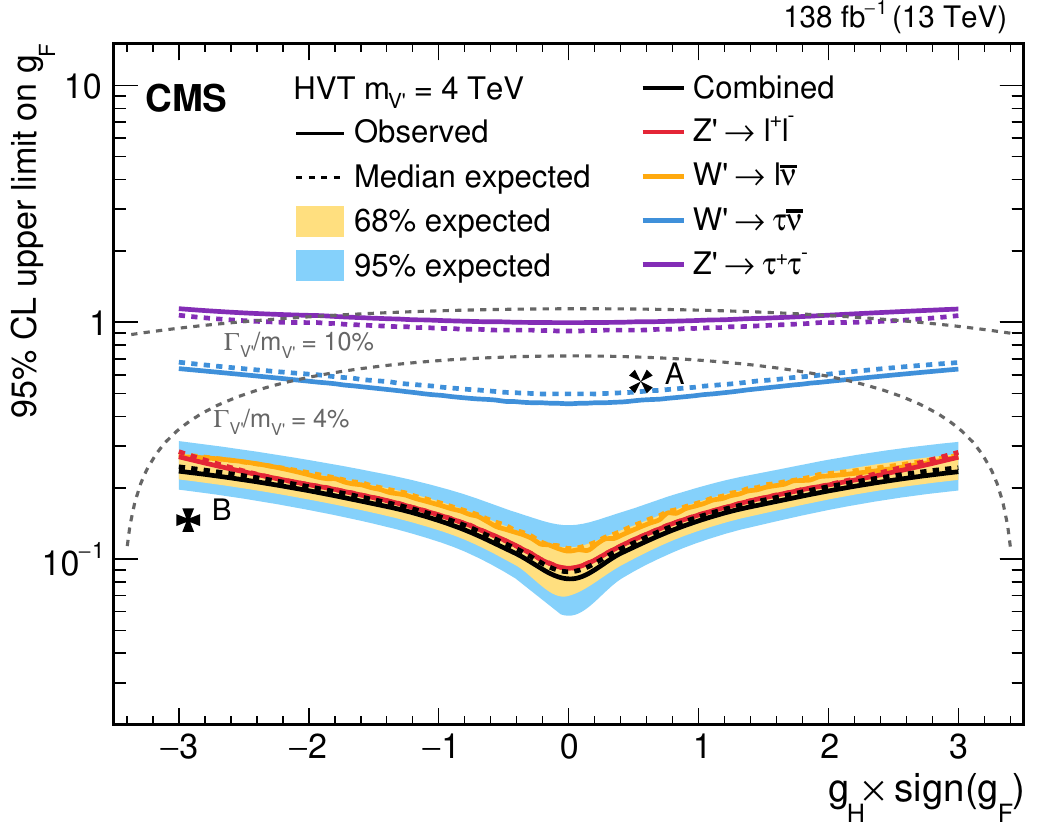}\\
\includegraphics[width=0.49\textwidth]{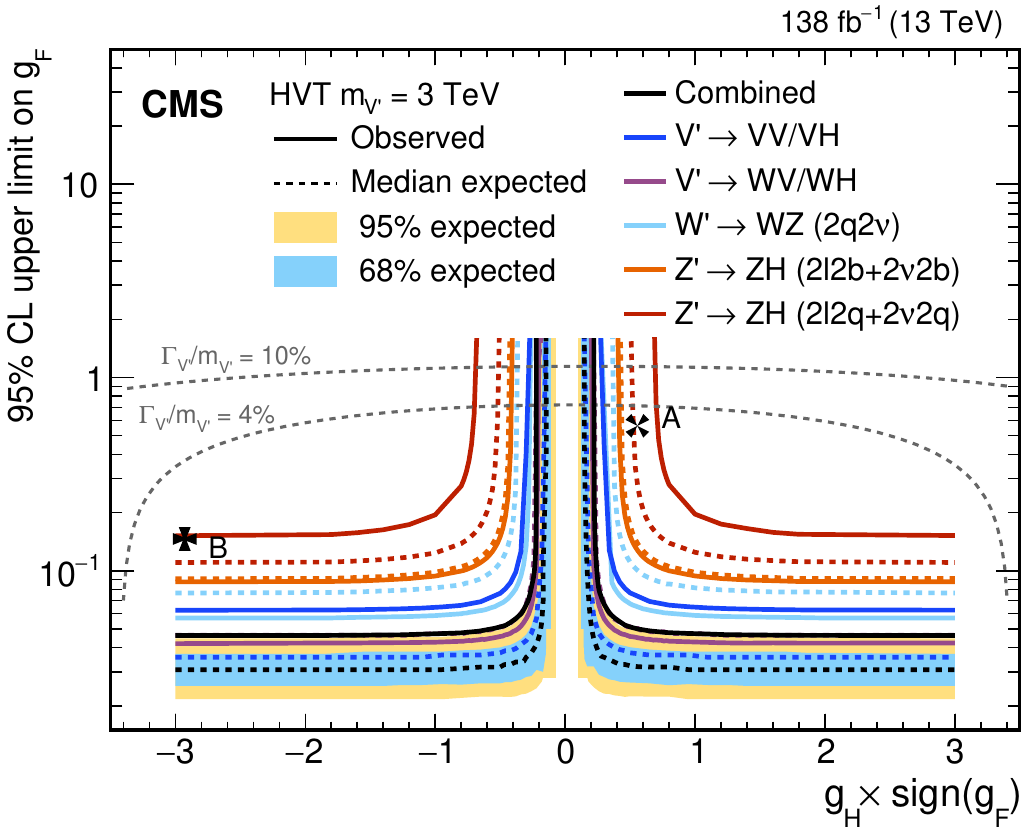}
\includegraphics[width=0.49\textwidth]{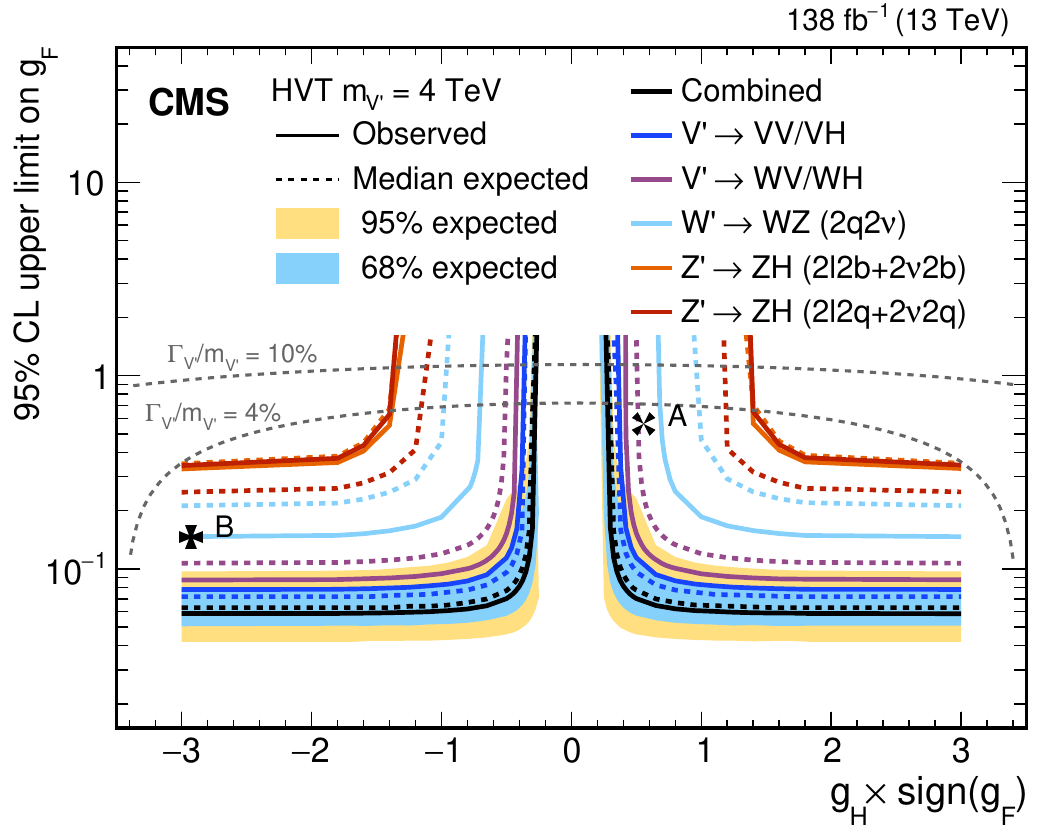}
\caption{Expected and observed 95\% \CL upper limits on the coupling parameter \gF as functions of \gH, under the assumption $\gV = 1$.
The results are shown separately for the \Vprimequarks (upper), \Vprimeleptons (middle), and \Vprimebosons (lower) categories for resonance masses of 3\TeV (left) and 4\TeV (right).
The results from the individual channels are also shown.
The gray dotted lines denote coupling values above which the relative width of the resonance, $\Gamma_{\Vprime}/\mV$,
exceeds 4 and 10\%. The couplings corresponding to the HVT models A and B are indicated by crosses.
}
\label{fig:2d_gf_gh}
\end{figure}

\begin{figure}[th]
\centering
\includegraphics[width=0.49\textwidth]{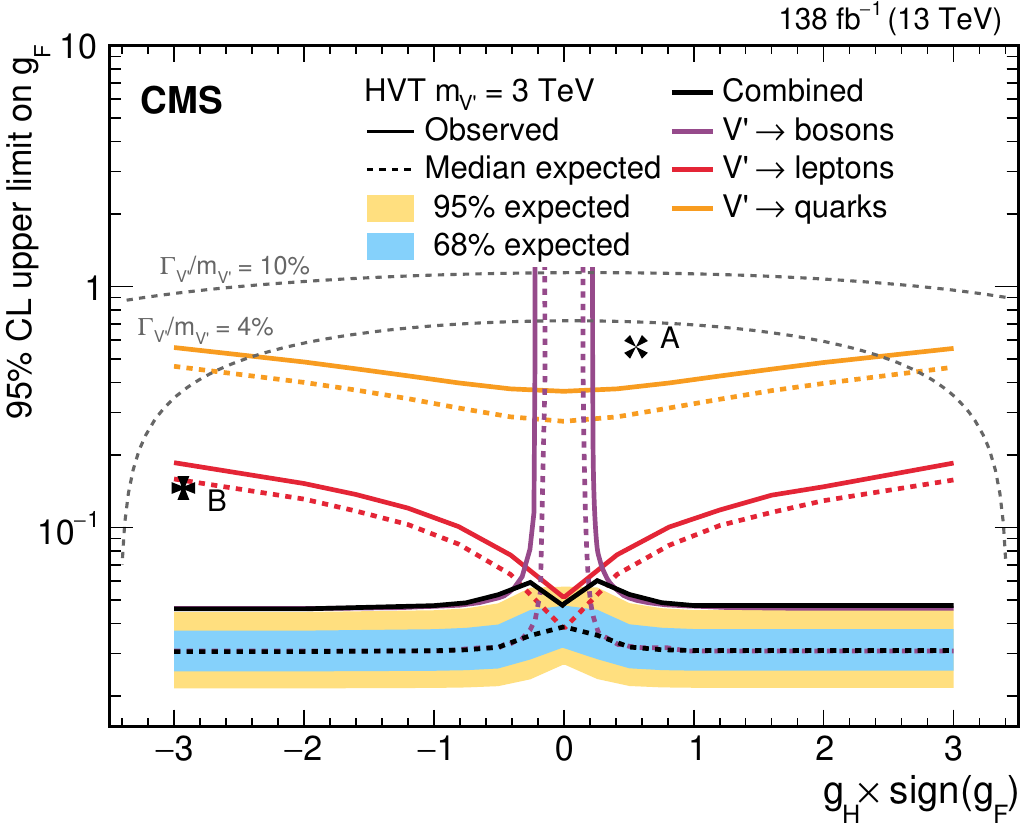}
\includegraphics[width=0.49\textwidth]{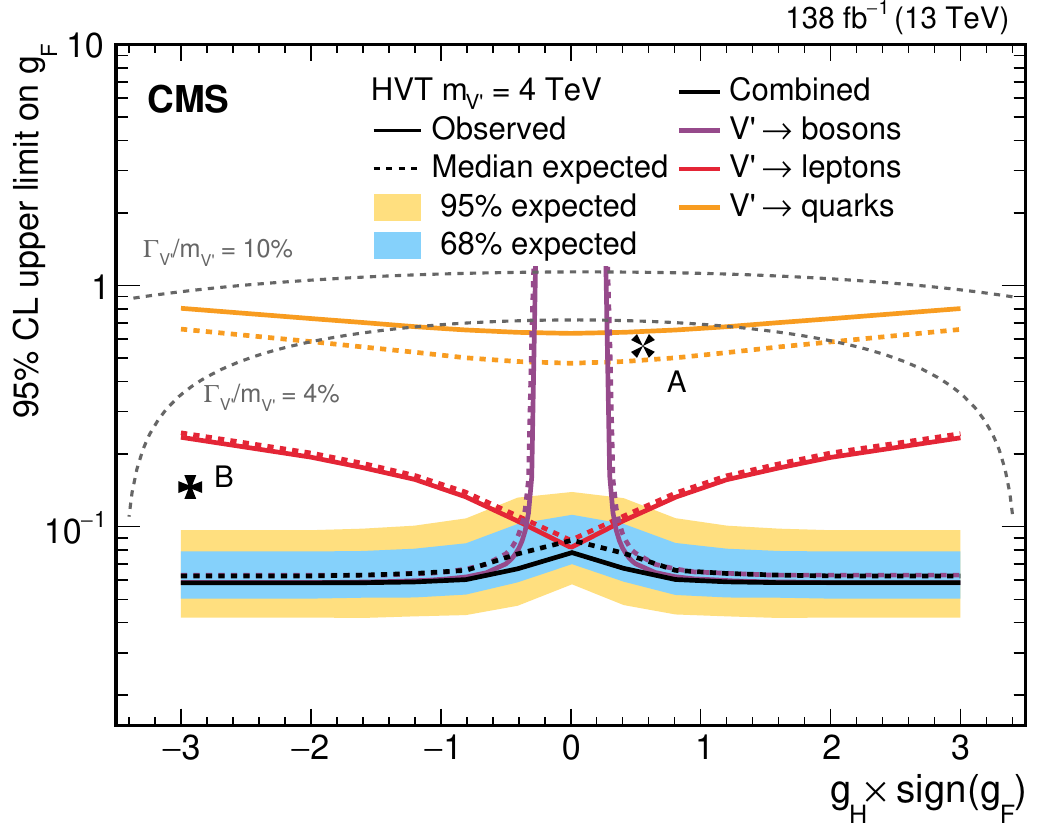}
\caption{Expected and observed 95\% \CL upper limits on the coupling parameter \gF as functions of \gH, under the assumption $\gV = 1$.
The combined results from all channels are shown for resonance masses of 3\TeV (left) and 4\TeV (right).
The results from the individual channels are also shown.
The gray dotted lines denote coupling values above which the relative width of the resonance, $\Gamma_{\Vprime}/\mV$,
exceeds 4 and 10\%. The couplings corresponding to the HVT models A and B are indicated by crosses.
}
\label{fig:2d_gf_gh_all}
\end{figure}

Additionally, exclusion limits are set in the HVT \gF-\gH coupling parameter space for representative resonance masses of 3 and 4\TeV, with $\mu$ and \gV fixed to 1.
The 3 and 4\TeV mass points are chosen as commonly used HVT benchmark values, enabling direct comparison with previous CMS and other LHC results.
The expected and observed 95\% \CL upper limits are shown in Figs.~\ref{fig:2d_gf_gh} and~\ref{fig:2d_gf_gh_all}, with results presented separately for the \Vprimequarks, \Vprimeleptons, and \Vprimebosons categories, as well as for their combination.
The same qualitative conclusions discussed for Figs.~\ref{fig:1d_individual_AB} and \ref{fig:1d_individual_C} hold here.
The \Vprimequarks category is dominated by the fully hadronic \Vprimeqq channel, the \Vprimeleptons category by the \Wprimelnu and \Zprimell channels, and the \Vprimebosons category by the fully hadronic \VV and \VH final states.
The sensitivity to the couplings is complementary across the channels.
At large values of \gH, the sensitivity is primarily driven by the \Vprimebosons category, while the \Vprimeleptons category dominates at small \gH values.
The full combination results for the \gF-\gH plane for resonance masses of 3 and 4\TeV are shown in Fig.~\ref{fig:2d_gf_gh_all}.
Compared to previous results with 2016 data alone, the constraint on \gF is improved by a factor of 1.9 to 2.6, depending on the resonance mass and the value of \gH.

To explore scenarios beyond the assumption of fermion universality,
the coupling strength \gF is split into four distinct components: \gql for light quarks, \gqh for third-generation quarks, \gll for light leptons, and \glh for third-generation leptons.
Together with $\mu$, \gV, and \gH, this extension totals seven POIs in the HVT framework.
For this study, the signal strength $\mu$ and the overall coupling \gV are fixed to 1, leaving five POIs.
We focus on two-dimensional parameter scans involving \gqh, since the largest sensitivity gains are observed in channels with third-generation quarks, such as \ttbar and \tb.
This leads to three key coupling planes:  \gqh-\gql, \gqh-\glh, and \gqh-\gH.

\begin{figure}[!ht]
\centering
\includegraphics[width=0.49\textwidth]{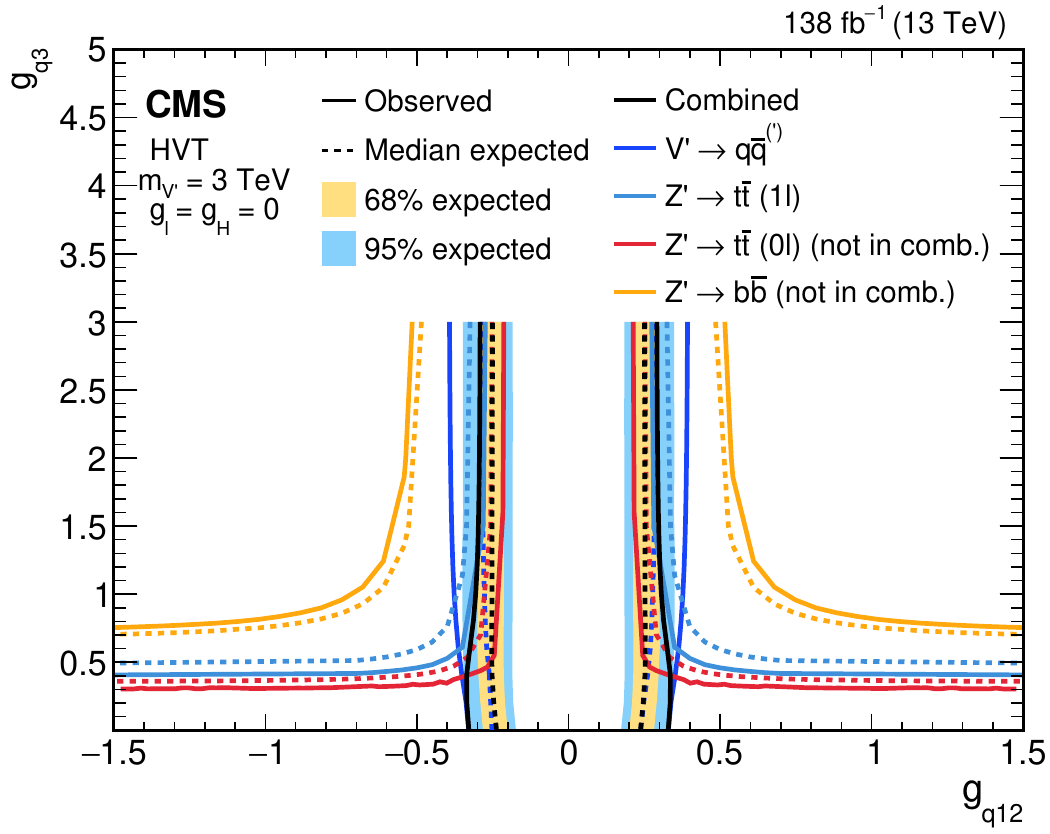}
\includegraphics[width=0.49\textwidth]{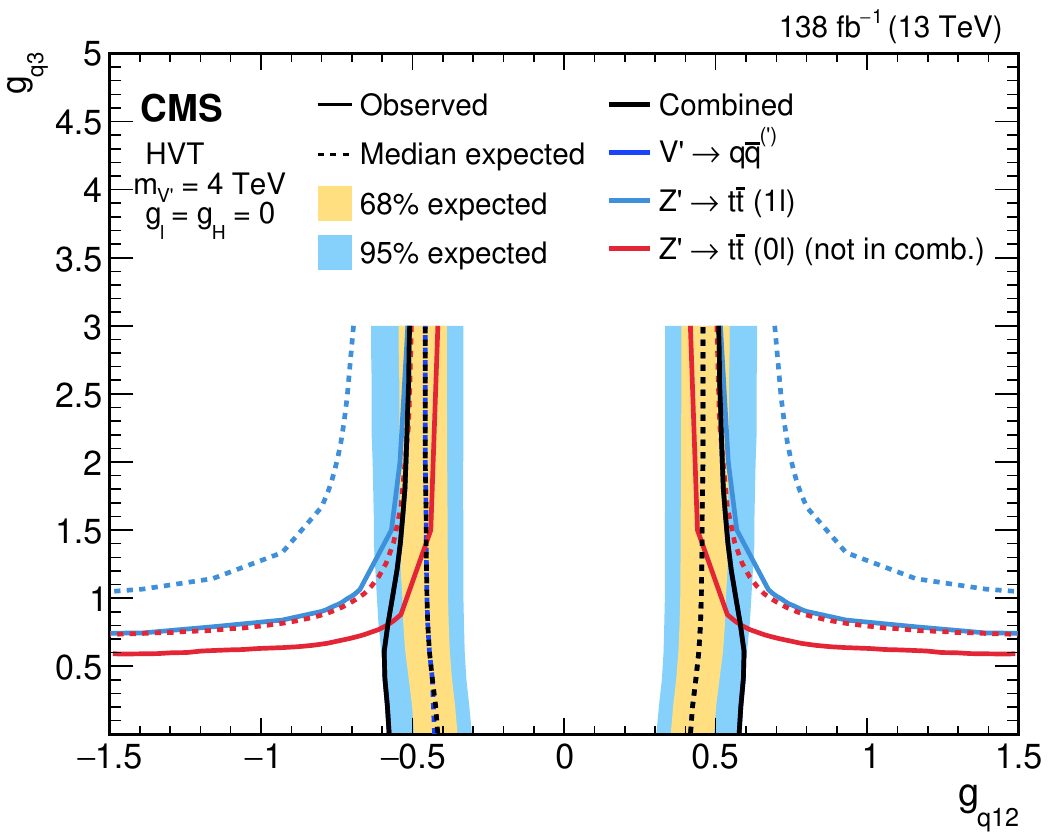}
\caption{Expected and observed 95\% \CL upper limits on the coupling parameter \gql as functions of \gqh,
under the assumption $\gV = 1$, $\gll = 0$, $\glh = 0$, and $\gH = 0$.
The results are shown for the \Vprimequarks category for resonance masses of 3\TeV (left) and 4\TeV (right).
The contributions from individual channels are also shown.
The line corresponding to the \Vprimeqq channel almost entirely overlaps with the combined result in the right plot.
Regions with large absolute values of \gql are excluded for any value of \gqh.
}
\label{fig:2d_gq3_gq12}
\end{figure}

In the first case, exclusion limits are evaluated in the \gqh-\gql plane, with \gll, \glh, and \gH set to zero.
This probes the possibility of nonuniversal couplings between light and third-generation quarks.
In this plane, the most sensitive channels are the \Zprimett channel at large \gqh values and the \Vprimeqq channel at large \gql values, as shown in Fig.~\ref{fig:2d_gq3_gq12}.

Additional scans are performed in the \gqh-\glh and \gqh-\gH planes, this time fixing the remaining couplings to their HVT model~A values.
Here, setting all other couplings to zero is not viable; for instance, setting $\gql = 0$ would eliminate DY production entirely.
The resulting parameter regions are fully excluded at 95\% CL, driven by the combined sensitivity of \Vprimequarks, \Vprimeleptons, and \Vprimebosons categories.
The complementarity of the individual channels allows for the complete exclusion of these parameter spaces, which could not be achieved by any single channel alone.

\section{Summary} \label{sec:Summary}

A combined statistical analysis of searches for heavy vector boson resonances decaying into pairs of \PW, \PZ, or \PH bosons, as well as into quark pairs (\qqbar, \bbbar, \ttbar, or \tb) or lepton pairs (\ellell, \lnu), with $\Pell = \Pe, \PGm, \PGt$, has been presented.
The results are based on proton-proton collision data at a center-of-mass energy of 13\TeV, corresponding to an integrated luminosity of 138\fbinv, collected by the CMS experiment from 2016 to 2018. No significant deviation from the expectations of the standard model is observed. The results are interpreted in the simplified heavy vector triplet framework, setting 95\% confidence level upper limits on the production cross sections of heavy vector bosons. These limits are further interpreted as constraints on their couplings to standard model particles.
The results exclude heavy vector boson resonances with a mass below 5.5\TeV in a weakly coupled scenario, below 4.8\TeV in a strongly coupled scenario, and up to 2.0\TeV in the case of production via vector boson fusion.
These results represent a significant improvement over previous constraints, driven by the inclusion of a larger data set and advances in the underlying analyses. Compared to results with the 2016 data alone, the sensitivity is improved by a factor of 1.7 to 3.6 for the strongly coupled scenario and 4.4 to 7.7 for the weakly coupled scenario. This combination provides the most stringent and comprehensive constraints to date on the production of heavy vector triplet resonances.

\begin{acknowledgments}
We congratulate our colleagues in the CERN accelerator departments for the excellent performance of the LHC and thank the technical and administrative staffs at CERN and at other CMS institutes for their contributions to the success of the CMS effort. In addition, we gratefully acknowledge the computing centers and personnel of the Worldwide LHC Computing Grid and other centers for delivering so effectively the computing infrastructure essential to our analyses. Finally, we acknowledge the enduring support for the construction and operation of the LHC, the CMS detector, and the supporting computing infrastructure provided by the following funding agencies: SC (Armenia), BMBWF and FWF (Austria); FNRS and FWO (Belgium); CNPq, CAPES, FAPERJ, FAPERGS, and FAPESP (Brazil); MES and BNSF (Bulgaria); CERN; CAS, MoST, and NSFC (China); MINCIENCIAS (Colombia); MSES and CSF (Croatia); RIF (Cyprus); SENESCYT (Ecuador); ERC PRG, TARISTU24-TK10 and MoER TK202 (Estonia); Academy of Finland, MEC, and HIP (Finland); CEA and CNRS/IN2P3 (France); SRNSF (Georgia); BMFTR, DFG, and HGF (Germany); GSRI (Greece); NKFIH (Hungary); DAE and DST (India); IPM (Iran); SFI (Ireland); INFN (Italy); MSIT and NRF (Republic of Korea); MES (Latvia); LMTLT (Lithuania); MOE and UM (Malaysia); BUAP, CINVESTAV, CONACYT, LNS, SEP, and UASLP-FAI (Mexico); MOS (Montenegro); MBIE (New Zealand); PAEC (Pakistan); MES, NSC, and NAWA (Poland); FCT (Portugal); MESTD (Serbia); MICIU/AEI and PCTI (Spain); MOSTR (Sri Lanka); Swiss Funding Agencies (Switzerland); MST (Taipei); MHESI (Thailand); TUBITAK and TENMAK (T\"{u}rkiye); NASU (Ukraine); STFC (United Kingdom); DOE and NSF (USA).

\hyphenation{Rachada-pisek} Individuals have received support from the Marie-Curie program and the European Research Council and Horizon 2020 Grant, contract Nos.\ 675440, 724704, 752730, 758316, 765710, 824093, 101115353, 101002207, 101001205, and COST Action CA16108 (European Union); the Leventis Foundation; the Alfred P.\ Sloan Foundation; the Alexander von Humboldt Foundation; the Science Committee, project no. 22rl-037 (Armenia); the Fonds pour la Formation \`a la Recherche dans l'Industrie et dans l'Agriculture (FRIA) and Fonds voor Wetenschappelijk Onderzoek contract No. 1228724N (Belgium); the Beijing Municipal Science \& Technology Commission, No. Z191100007219010, the Fundamental Research Funds for the Central Universities, the Ministry of Science and Technology of China under Grant No. 2023YFA1605804, the Natural Science Foundation of China under Grant No. 12535004, and USTC Research Funds of the Double First-Class Initiative No.\ YD2030002017 (China); the Ministry of Education, Youth and Sports (MEYS) of the Czech Republic; the Shota Rustaveli National Science Foundation, grant FR-22-985 (Georgia); the Deutsche Forschungsgemeinschaft (DFG), among others, under Germany's Excellence Strategy -- EXC 2121 ``Quantum Universe" -- 390833306, and under project number 400140256 - GRK2497; the Hellenic Foundation for Research and Innovation (HFRI), Project Number 2288 (Greece); the Hungarian Academy of Sciences, the New National Excellence Program - \'UNKP, the NKFIH research grants K 131991, K 133046, K 138136, K 143460, K 143477, K 146913, K 146914, K 147048, 2020-2.2.1-ED-2021-00181, TKP2021-NKTA-64, and 2025-1.1.5-NEMZ\_KI-2025-00004 (Hungary); the Council of Science and Industrial Research, India; ICSC -- National Research Center for High Performance Computing, Big Data and Quantum Computing, FAIR -- Future Artificial Intelligence Research, and CUP I53D23001070006 (Mission 4 Component 1), funded by the NextGenerationEU program (Italy); the Latvian Council of Science; the Ministry of Education and Science, project no. 2022/WK/14, and the National Science Center, contracts Opus 2021/41/B/ST2/01369, 2021/43/B/ST2/01552, 2023/49/B/ST2/03273, and the NAWA contract BPN/PPO/2021/1/00011 (Poland); the Funda\c{c}\~ao para a Ci\^encia e a Tecnologia, grant CEECIND/01334/2018 (Portugal); the National Priorities Research Program by Qatar National Research Fund;  MICIU/AEI/10.13039/501100011033, ERDF/EU, ``European Union NextGenerationEU/PRTR", and Programa Severo Ochoa del Principado de Asturias (Spain); the Chulalongkorn Academic into Its 2nd Century Project Advancement Project, the National Science, Research and Innovation Fund program IND\_FF\_68\_369\_2300\_097, and the Program Management Unit for Human Resources \& Institutional Development, Research and Innovation, grant B39G680009 (Thailand); the Kavli Foundation; the Nvidia Corporation; the SuperMicro Corporation; the Welch Foundation, contract C-1845; and the Weston Havens Foundation (USA).
\end{acknowledgments}

\bibliography{auto_generated}
\appendix
\numberwithin{table}{section}
\numberwithin{figure}{section}

\section{Additional figures}
\label{app:suppMat}

\begin{figure}[th]
\centering
\includegraphics[width=0.6\textwidth]{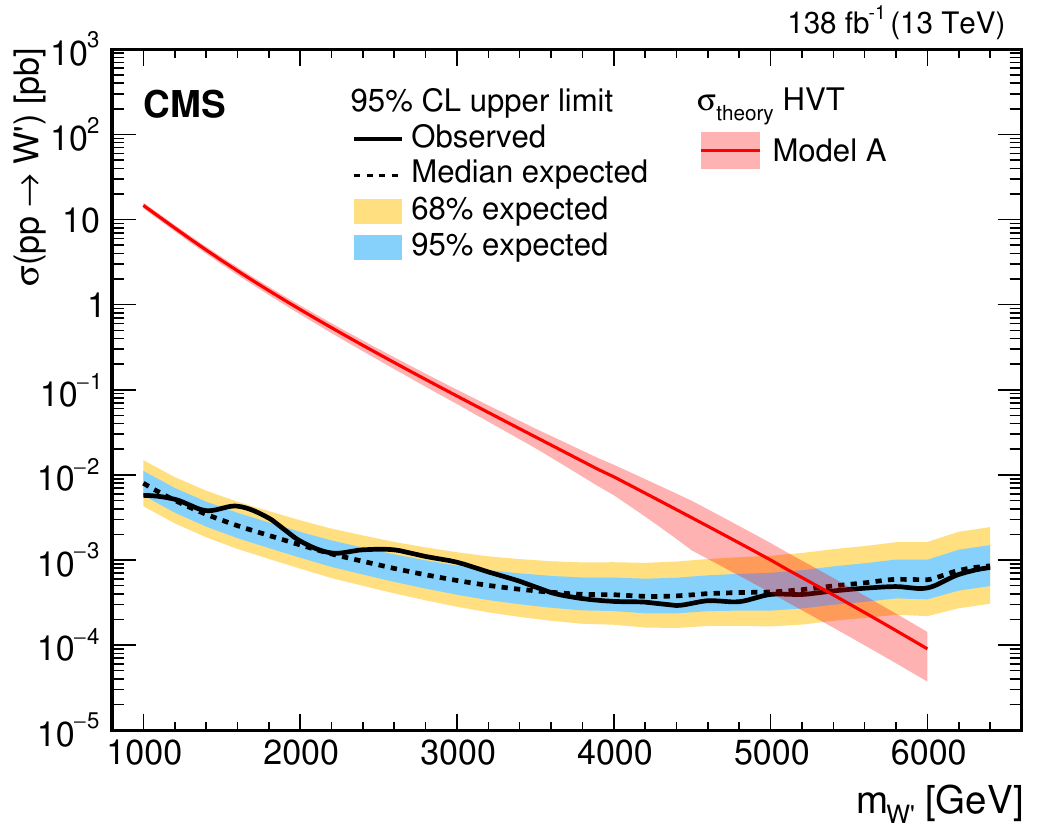}
\includegraphics[width=0.6\textwidth]{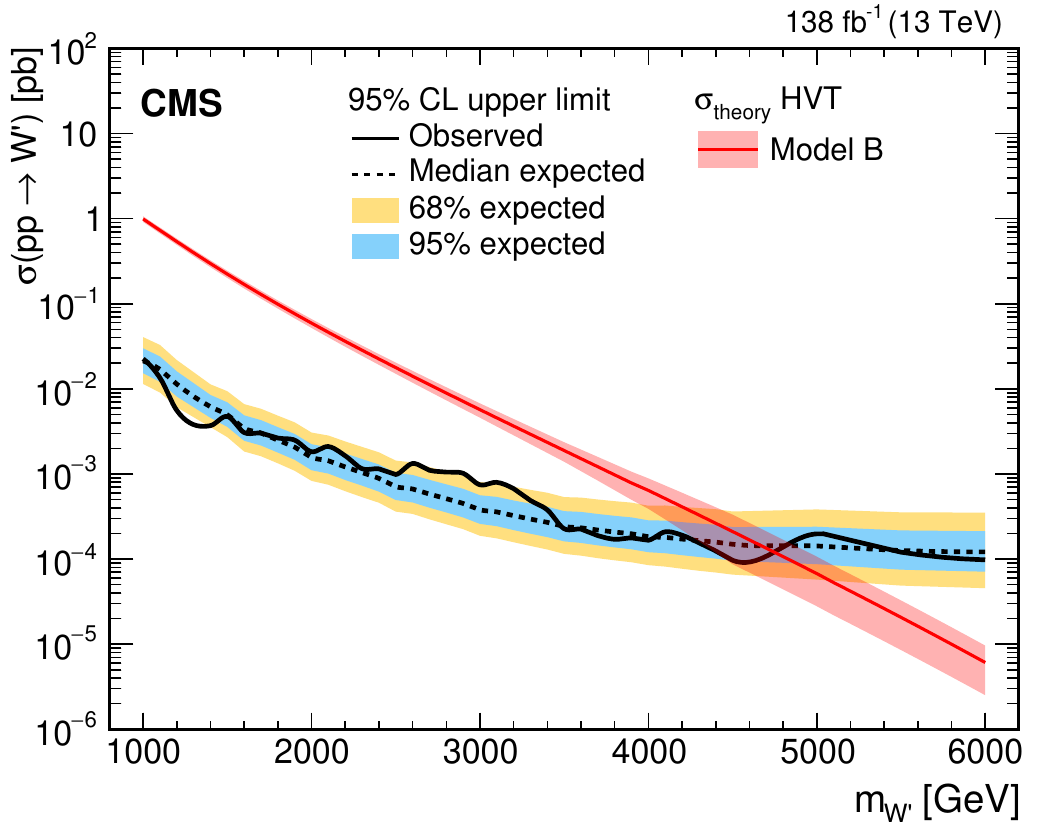}
\includegraphics[width=0.6\textwidth]{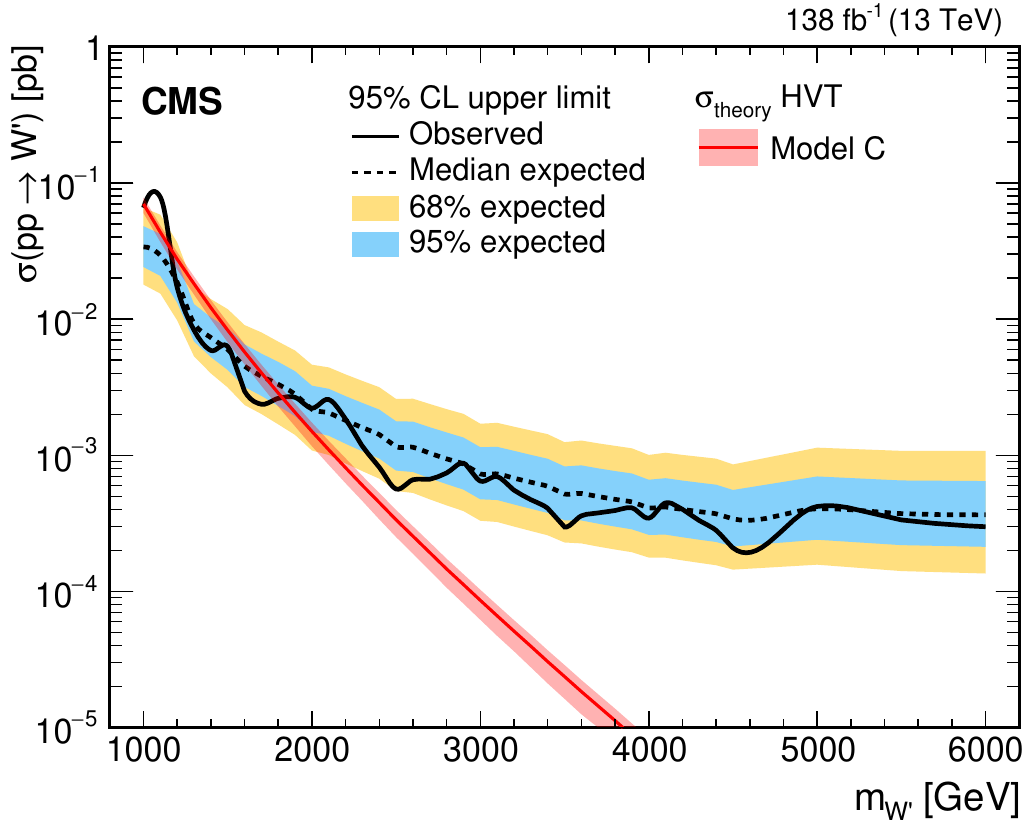}
\caption{Expected and observed 95\% \CL upper limits on the \Wprime boson production cross section
as a function of the resonance mass \mWpr for the \Wprime interpretations.
The limits are evaluated in the HVT model~A (upper), B (middle), and C (lower) scenarios.}
\label{fig:1d_combined_W_sep}
\end{figure}

\begin{figure}[th]
\centering
\includegraphics[width=0.6\textwidth]{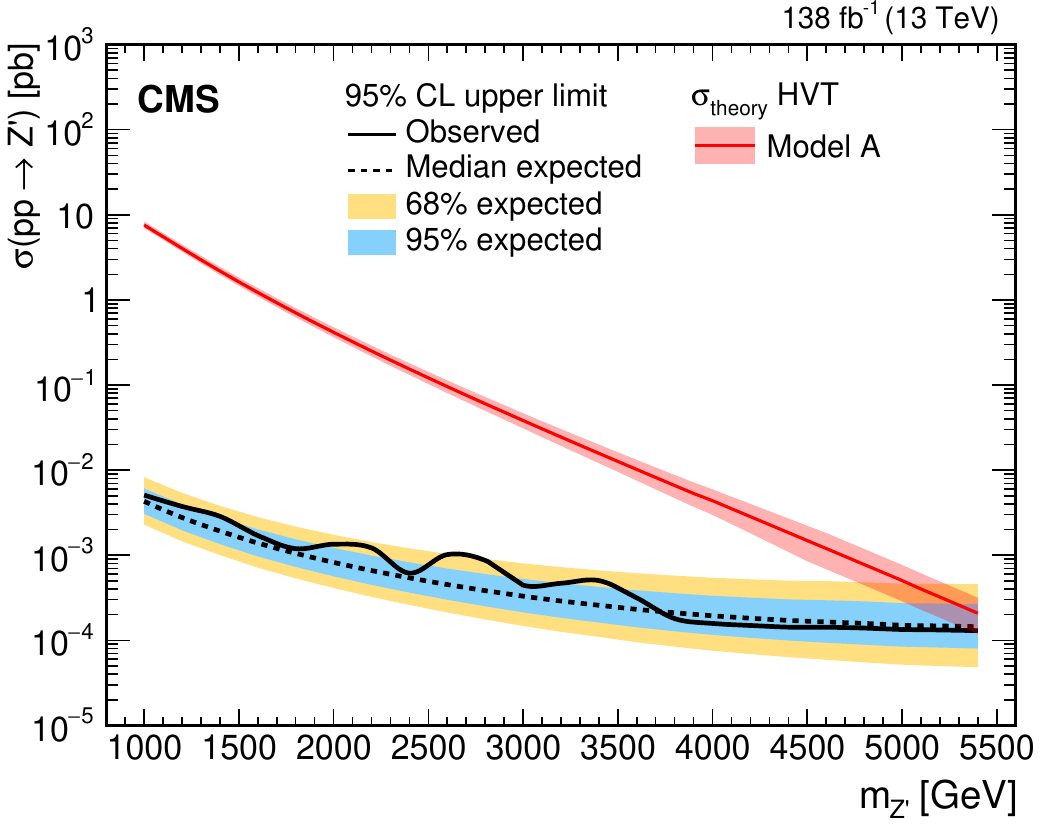}
\includegraphics[width=0.6\textwidth]{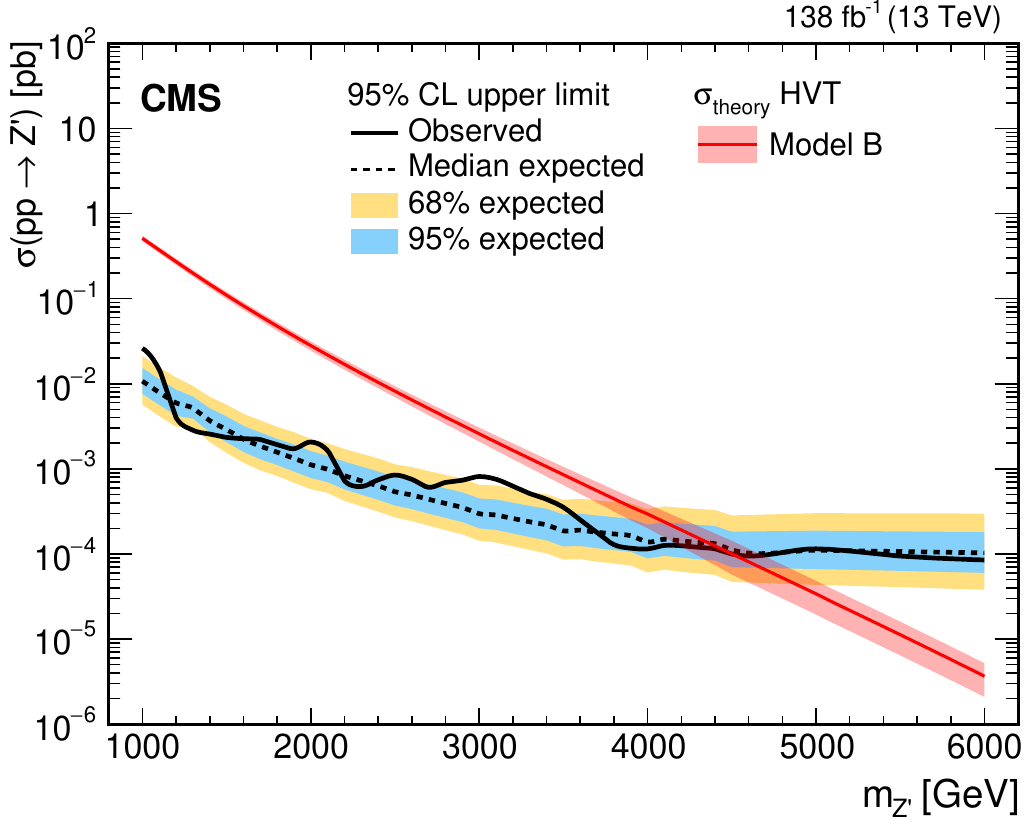}
\includegraphics[width=0.6\textwidth]{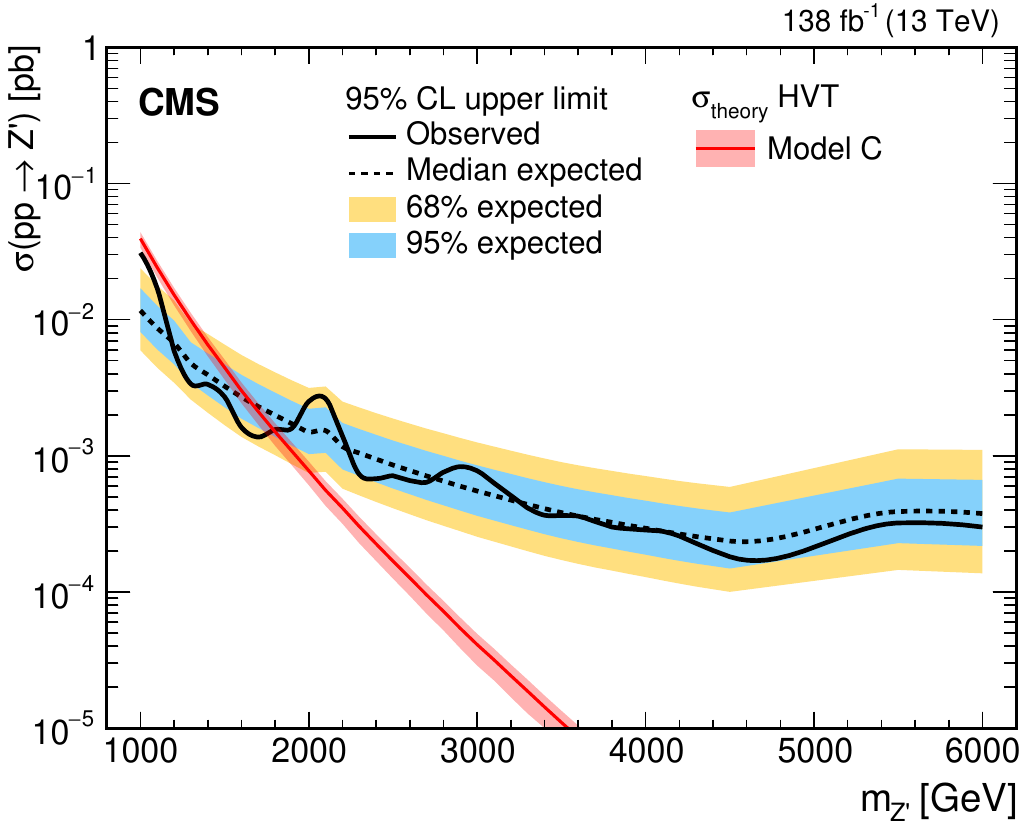}
\caption{Expected and observed 95\% \CL upper limits on the \Zprime boson production cross section
as a function of the resonance mass \mZpr for the \Zprime interpretations.
The limits are evaluated in the HVT model~A (upper), B (middle), and C (lower) scenarios.}
\label{fig:1d_combined_Z_sep}
\end{figure}

A breakdown of the exclusion limits by \Wprime and \Zprime combinations under the three HVT
benchmark models is shown in Figs.~\ref{fig:1d_combined_W_sep} and \ref{fig:1d_combined_Z_sep}.

\cleardoublepage \section{The CMS Collaboration \label{app:collab}}\begin{sloppypar}\hyphenpenalty=5000\widowpenalty=500\clubpenalty=5000\cmsinstitute{Yerevan Physics Institute, Yerevan, Armenia}
{\tolerance=6000
A.~Hayrapetyan, V.~Makarenko\cmsorcid{0000-0002-8406-8605}, A.~Tumasyan\cmsAuthorMark{1}\cmsorcid{0009-0000-0684-6742}
\par}
\cmsinstitute{Institut f\"{u}r Hochenergiephysik, Vienna, Austria}
{\tolerance=6000
W.~Adam\cmsorcid{0000-0001-9099-4341}, L.~Benato\cmsorcid{0000-0001-5135-7489}, T.~Bergauer\cmsorcid{0000-0002-5786-0293}, M.~Dragicevic\cmsorcid{0000-0003-1967-6783}, C.~Giordano, P.S.~Hussain\cmsorcid{0000-0002-4825-5278}, M.~Jeitler\cmsAuthorMark{2}\cmsorcid{0000-0002-5141-9560}, N.~Krammer\cmsorcid{0000-0002-0548-0985}, A.~Li\cmsorcid{0000-0002-4547-116X}, D.~Liko\cmsorcid{0000-0002-3380-473X}, M.~Matthewman, J.~Schieck\cmsAuthorMark{2}\cmsorcid{0000-0002-1058-8093}, R.~Sch\"{o}fbeck\cmsAuthorMark{2}\cmsorcid{0000-0002-2332-8784}, M.~Shooshtari\cmsorcid{0009-0004-8882-4887}, M.~Sonawane\cmsorcid{0000-0003-0510-7010}, W.~Waltenberger\cmsorcid{0000-0002-6215-7228}, C.-E.~Wulz\cmsAuthorMark{2}\cmsorcid{0000-0001-9226-5812}
% ...
\par}
\vskip\cmsinstskip
\dag:~Deceased\\
$^{1}$Also at Yerevan State University, Yerevan, Armenia\\
$^{2}$Also at TU Wien, Vienna, Austria\\
% ...
\end{sloppypar}
%%% END EDITABLE REGION %%%
% skeleton_end
\end{document}